\documentclass[fleqn,usenatbib]{mnras}

\usepackage{newtxtext,newtxmath}

\usepackage[T1]{fontenc}
\usepackage{ae,aecompl}

\usepackage{graphicx}	
\usepackage{amsmath}	
\usepackage{amssymb}	
\usepackage{threeparttable}
\usepackage{color}

\title[Self-gravity and radiative cooling]{The effects of disk self-gravity and radiative cooling on the formation of gaps and spirals by young planets}

\author[S. Zhang \& Z. Zhu]{
Shangjia Zhang,$^{1}$\thanks{E-mail: shangjia.zhang@unlv.edu}
Zhaohuan Zhu$^{1}$
\\
$^{1}$Department of Physics and Astronomy, University of Nevada, Las Vegas, 4505 S. Maryland Pkwy, Las Vegas, NV, 89154, USA\\
}

\pubyear{2019}

\begin{document}
\label{firstpage}
\pagerange{\pageref{firstpage}--\pageref{lastpage}}
\maketitle

\begin{abstract}
We have carried out two-dimensional hydrodynamical simulations to study the effects of disk self-gravity and radiative cooling on the formation of gaps and spirals. (1) With disk self-gravity included, we find stronger, more tightly-wound spirals and deeper gaps in more massive disks. The deeper gaps are due to the larger Angular Momentum Flux (AMF) of the waves excited in more massive disks, as expected from the linear theory. The position of the secondary gap does not change, provided that the disk is not extremely massive ($Q \gtrsim 2$).  (2) With radiative cooling included, the excited spirals become monotonically more open (less tightly-wound) as the disk's cooling timescale increases. On the other hand, the amplitude and strength of the spirals decrease when the cooling time increases from a small value to $\sim 1/\Omega$,  but then the amplitude starts to increase again when the cooling time continues to increase. This indicates that radiative dissipation becomes important for waves with $T_{cool}\sim$ 1. Consequently, the induced primary gap is narrower and the secondary gap becomes significantly shallower when the cooling time becomes $\sim 1/\Omega$. When the secondary gap is present, the position of it moves to the inner disk from the fast cooling cases to the slow cooling cases. The dependence of gap properties on the cooling timescale (e.g., in AS 209) provides a new way to constrain the disk optical depth and thus disk surface density. 
\end{abstract}




\section{Introduction}\label{sec:intro}
Protoplanetary disks from ALMA observations reveal many substructures (e.g., gaps, rings, and spiral arms) in dust continuum emission \citep{andrews18b, long18, perez16}. Among them, the most common features are concentric gaps and rings \citep{huang18b}.
There are many interpretations for these features, such as zonal flows \citep{flock15}, aggregate sintering \citep{okuzumi2016}, secular gravitational instabilities \citep{takahashi2014}, self-induced dust pile-ups \citep{gonzalez2017},  dust growth at snowlines \citep{zhang2015,pinilla2017},  planet-disk interactions \citep{Ovelar2013,dong2015a}, and so on. Recent high-resolution observations for a relatively large sample of disks \citep{huang18b, long18, vandermarel19} conclude that snowlines cannot consistently match the positions of these gap features, but it may still apply to individual objects. 

Of all the exciting scenarios, this paper focuses on the interpretation of planet-disk interactions.
Assuming these features are due to the planets, many authors infer the planet mass from the gap properties by comparing observations with hydrodynamical simulations. These properties include gap width and depth in dust emission \citep{zhang18}, and sub/super keplerian rotational velocity in CO channel maps \citep{teague18a, zhang18, han19}. Most past works associate each gap with a planet. Recent simulations show more promising results that one single planet can explain up to five gaps in ALMA observations \citep{bae2017, guzman18,zhang18}.  These multiple gaps that are induced by a single planet is due to the presence of multiple spirals, which results from constructive interference of the density waves at different $m$ modes in the inner disk \citep{bae2018a, bae2018b, miranda19a}. The one-armed spiral opens the primary (major) gap at the position of the planet, and the secondary spiral opens the secondary gap in the inner disk and so on. The spacing of the gaps is also useful to constrain the gaseous disk scale height $h/r$  since the ratio of the positions between the secondary gap and the primary gap mainly depends on the disk $h/r$ \citep{dong18b, zhang18}. If we can further constrain $h/r$ using other observables like gap widths and depths, we can use the position of the secondary gap to even constrain the planet mass \citep{kanagawa_depth, kanagawa_width, bae2017,  zhang18, han19}.

However, most simulations that are used to infer the planet mass from the disk substructures neglect the (1) self-gravity from the disk and (2) radiative cooling. For the latter, they employ locally isothermal Equation of State (EoS) instead, which is equivalent to instant cooling. These two simplifications might not be valid to realistic disks. Consequently, the planet masses inferred from observations might be subject to systematic errors. Self-gravity and radiative cooling can play essential roles in some disks.

Several recent works suggest disks might be more massive than previously thought. \citet{booth19} measure the HD 163296 disk mass using $^{13}$C$^{17}$O line, and find the gas mass is a factor of 2-6 higher than previously estimated using C$^{18}$O.  Using the DSHARP opacity \citep{birnstiel18} and optical-thin assumption,  \citet{zhang18} find that Toomre $Q$ at the gap edges are $\lesssim$ 10 for most of the disks (see Table 3 therein). Among the same series of papers,  \citet{dullemond18b} also find that most of the prominent rings are marginally gravitationally stable if the gas-to-dust ratio is 100 (see Figure 7 therein). Furthermore,  \citet{zhu19b} suggests that these disks might be very optically thick, even though dust scattering makes these disks look like optically thin. Without the assumption of the dust opacity, \citet{powell19} analyze seven disks by measuring the locations of ``dust lines''---the cutoff radii of continuum emission---at different wavelengths, and also find all of those disks have $Q \lesssim 10$, with several approaching unity. 
 For these reasons, disk self-gravity can become quite important for the interactions between planets and disks.

The disk should also have orders of magnitude difference in cooling time at different radii \citep{zhu2015, miranda19b}. Assuming the Minimum Mass Solar Nebula (MMSN), the cooling is faster at the outer disk and slower at the inner disk ---  7 orders of magnitude of the cooling time between 1 au ($10^{5} /\Omega$) and 100 au ($10^{-2} /\Omega$) \citep{zhu2015}. The fast cooling resembles the locally isothermal disk, whereas the slow cooling represents the adiabatic disk. However, it is unclear how the disk might look like beyond these two extremes. Radiative cooling with different cooling timescales at different locations in disks might further change the disk substructures.

In this paper, we study spirals and gaps induced by planets in disks with non-negligible self-gravity and radiative cooling. In Section \ref{sec:theory}, we lay out the theoretical background of the linear theory and some basic quantities being used throughout the paper. In Section \ref{sec:method}, we introduce the simulation setups. In Section \ref{sec:results}, we present the results after adding these two physical processes separately, and also explore a situation that includes both self-gravity and radiative cooling. We also compare our simulation results with previous analytical results from  \citet{goldreich80} and  \citet{miranda19a}. In Section \ref{sec:discussion}, we quantify the change of AMF and discuss the observational implications in linear regime. Then, we take AS 209 disk as a test bed for these two processes, and address some limitations of our simulations. Finally, we conclude this paper in Section \ref{sec:conclusion}.

\section{Theoretical Background}\label{sec:theory}
 \citet{goldreich78, goldreich79, goldreich80} develop the linear theory for planet-disk interactions. This theory produces insightful analytical results which can be used to understand numerical simulations on planet migration and gap opening.

In order to study planet-disk interactions in the linear regime, the planet mass should be less than the thermal mass \citep{goodman2001}
\begin{equation}
    M_{th} = \Big(\frac{h}{r}\Big)_p^3 M_*  = 1 M_J \Big(\frac{(h/r)_p}{0.1}\Big)^3 \frac{M_*}{M_\odot}  \label{eq:Mth}\,,
\end{equation}
where $(h/r)_p$ is the disk's aspect ratio at the planet's position. $M_*$ is the mass of the central star. At the thermal mass, the gaseous disk scale height is comparable to  both the Hill and Bondi radii of the planet, and the distance between the planet and spiral shock forming region is comparable to the planet's Bondi radius so that spiral shocks begin to affect planet accretion \citep{bethune2019a}. If $M_p$ $\gtrsim$ $M_{th}$, the density wave excited by the planet starts from the non-linear regime, so that the spiral waves immediately become spiral shocks after the wave excitation. Below the thermal mass, the spiral shock region can be well separated from the wave launching area.  Most of the simulations in this paper have $M_p \lesssim M_{th}$, but we also explore several cases in the high mass regime.

Angular momentum transport is one of the key aspects in the accretion theory.  In the linear theory, the changing rate of the angular momentum in the disk enclosed within a radius $r$ is equal to the planet's torque $\Gamma$ on that region minus the Angular Momentum Flux\footnote{This is coined as angular momentum current in  \citet{binney08}, denoted by capital ``C''.} (AMF) $F_J$ that flows out of the region, that is
\begin{equation}
    \frac{{dL}}{dt} = {\Gamma}(r) - F_J(r).
    \label{eq:dLdt}
\end{equation}

\subsection{Torque}
The first term on the right-hand side of Equation \ref{eq:dLdt} is the planet's torque to the disk region within $r$,
\begin{equation}
    \Gamma = - \int_{disk} \Sigma(\vec{r} \times \vec{F}) df = \int_r \frac{dT}{dr} dr,
    \label{eq:torque}
\end{equation}
as in  \citet{kley12}, where
\begin{equation}
    \frac{dT}{dr}(r) = r\int_0^{2\pi}\Sigma(r,\phi)\frac{\partial \Phi_p}{\partial \phi} d\phi\label{eq:dTdr}
\end{equation}
is torque density and $\Phi_p$ is the planet's potential in the disk coordinate,
\begin{equation}
    \Phi_p = -\frac{GM_p}{\big\{r_p^2+r^2-2r_pr\mathrm{cos}(\phi-\phi_p)+s^2\big\}^{1/2}},
    \label{eq:planetpotential}
\end{equation}
where $r_p$, $\phi_p$ are planet's position in radial and azimuthal directions and $s$ is the smoothing length used in simulations to avoid the singularity of the planetary potential. The planet feels the opposite torque, which leads to planet migration if the torque from the inner disk is not balanced by the torque from the outer disk.

\citet{goldreich80} derive the one-sided torque as 
\begin{equation}
    F_{J0} = (M_p/M_*)^2h_p^{-3}\Sigma_p r_p^4\Omega_p^2
    \label{eq:FJ0}
\end{equation}
where $\Omega_p$ is the angular frequency of the planet (all the ``$p$'' as a subscript in the paper refers to that quantity evaluated at the position of the planet). This value is commonly used to normalize the torque and AMF.

\subsection{AMF} \label{sec:AMF}
The second term on the right-hand side of Equation \ref{eq:dLdt} is the AMF carried by the disturbance (e.g., spiral waves). The angular momentum carried by the spiral waves will be eventually deposited to the background disk to induce gaps in disks. In order to understand the gap properties, such as depth, width, and secondary gap positions, it is necessary to understand the  AMF of the spiral waves. With disk self-gravity included in the analysis, AMF has two components, advective AMF, $F_A$, and gravitational AMF, $F_G$ \citep{lyndenbell72, goldreich79}. The latter only occurs in self-gravitating disks.  The total AMF is, $F_J = F_A + F_G$. 

$F_A$ is the advective transport of angular momentum due to the  motion of the fluid and leads to a flow of angular momentum through circumference at radius $r$. It is also related to the Reynolds stress in turbulence studies. It can be calculated as,
\begin{equation}
    F_A(r) = r^2\Sigma(r)\oint u_r(r,\phi)u_{\phi}(r,\phi)d\phi ,
    \label{eq:AMF}
\end{equation}
where $\Sigma$ is the gaseous disk surface density and $u_r$ and $u_\phi$ are the velocity perturbations in the $r$ and $\phi$ directions.

$F_G$ is due to the non-axisymmetric part of the disk potential. The spiral structure produces a spiral gravitational field, which exerts torque and transfers angular momentum from one part of the disk to another. It is non-zero in self-gravitating disks and becomes important when the following Tommre $Q$ parameter is small,
\begin{equation}
    Q = \frac{c_s \Omega}{\pi G\Sigma},
    \label{eq:toomreQ}
\end{equation}
where $c_s$ is the sound speed and $c_s$ = $(h/r) v_\phi$. Given fixed $c_s$ and $\Omega$, a smaller $Q$ means a higher disk surface density. When $Q=1$, the disk becomes gravitationally unstable. The value of $F_G$ is the torque exerted to the inner disk, 
\begin{equation}
    F_G(r) =  \int_{0}^{r} dr' r' \int_{0}^{2\pi} d\phi \Sigma \frac{\partial\Phi_{out}}{\partial\phi},
    \label{eq:FG}
\end{equation}
where $\Phi_{out}$ is the disk potential at $r$ (the formula is the same as Equation \ref{eq:dTdr}, except that $\Phi_p$ is replaced with $\Phi_{out}$).
For trailing spiral arms, the angular momentum is transferred from the inner to the outer disk, since the inner disk exerts positive gravitational torque on the outer disk \citep{lyndenbell72,binney08}.  

Note that, in globally isothermal disks, the total AMF is conserved ($dF_J/dr$ = 0) based on the linear theory \citep{goldreich79}. However, as  \citet{miranda19b} point out, this is not the case in locally isothermal disks due to the torque applied onto the wave by the background shear flow (this is also reported in \citealt{lin2011, lin2015}). In locally isothermal disks, $d(F_J/c_s^2)/dr$ = 0 is conserved. 

In the linear theory, all quantities can be further Fourier decomposed into individual $m$ harmonics,
\begin{equation}
    \eta(r,\phi) = \sum_{m=-\infty}^{\infty} \eta_m(r)e^{im(\phi-\phi_p)}
    \label{eq:fourier}
\end{equation}
where $\eta$ can be $u_r$, $u_\phi$, and $\Sigma$, and $\eta_m$ is a complex number. Putting $u_r(r)$ and $u_\phi(r)$ in Equation \ref{eq:AMF},
\begin{equation}
   \begin{split}
   F_{A}(r) 
   &= 2\pi r^2\Sigma(r) \sum_{m=-\infty}^{\infty}u_{r,m}(r)u_{\phi,m}^{*}(r) \\
   &= 4\pi r^2\Sigma(r) \sum_{m=0}^{\infty} \Re[u_{r,m}(r)u_{\phi,m}^{*}(r)].
   \label{eq:AMFexpend}
   \end{split}
\end{equation}
The second equality holds because the Fourier transform of real signals is Hermitian.
If $F_A$ is expanded with positive integer harmonics (the $m=0$ term is always zero),
\begin{equation}
    F_{A}(r)= \sum_{m=1}^{\infty} F_{A,m}(r),
\end{equation}
each $m$ component of the advective AMF can be expressed as,
\begin{equation}
   \begin{split}
   & F_{A,m}(r) = 4\pi r^2\Sigma(r)  \Re[u_{r,m}(r)u_{\phi,m}^{*}(r)]  \\
    & = 4\pi r^2\Sigma(r)  \Big(\Re[u_{r,m}(r)]\Re[u_{\phi,m}(r)]+\Im[u_{r,m}(r)]\Im[u_{\phi,m}(r)]\Big),
   \label{eq:AMFm}
   \end{split}
\end{equation}
which is identical to the result using continuous Fourier transform in the shearing sheet geometry \citep{dong2011a, rafikov12}.
Likewise, the total AMF can be expanded as,
\begin{equation}
    F_{J}(r)= \sum_{m=1}^{\infty} F_{J,m}(r).
\end{equation}
The sum goes to infinity, but the normalized AMF ($F_{J,m}/F_{F_{J0}}$) reaches the maximum at $m \sim$ $(1/2)/(h/r)$ and becomes $\ll 1$ at several $1/(h/r)$. Thus, the sum can stop at some $m$ \citep{goldreich80,artymowicz93a,artymowicz93b,ward97}, which is known as the ``torque cutoff''. 

\subsubsection{Relative importance between $F_A$ and $F_G$}
Here we discuss the relative importance between $F_A$ and $F_G$. We borrow the WKB solutions in the tightly-wound limit ($k \gg 1/R$) obtained by \citet{goldreich79}. Their analytical solution for the advective AMF is, 
\begin{equation}
    F_A = -\frac{\pi m r\Sigma k}{2\pi G\Sigma |k| - k^2c_s^2}\Big(1-\frac{c^2 |k|}{2\pi G\Sigma}\Big)^2\Phi(r)^2.
\end{equation}
where $\Phi(r)$ is the amplitude of the perturbed disk gravitational potential in the form of WKB solution. The gravitational AMF is,
\begin{equation}
    F_G = \mathrm{sgn}(k)\frac{mr\Phi^2(r)}{4G}\,.
\end{equation}
Taking the ratio between the $F_A$ and the $F_G$, and define the critical wave number $k_{crit}$ \citep{binney08} as,
\begin{equation}
    k_{crit} = \frac{\Omega^2}{2\pi G \Sigma} = \frac{Q}{2}\frac{\Omega}{c_s} = \frac{Q}{2h},
    \label{eq:kcrit}
\end{equation}
we get,
\begin{equation}
    \frac{F_A}{F_G} = 2\Big(\frac{Q^2}{4}\frac{|k|}{k_{crit}} - 1\Big) = 2\Big(\frac{|k|}{k_{cross}} - 1\Big), \label{eq:FaFg}
\end{equation}
where  $k_{cross} = 4 k_{crit}/Q^2 = 2/Qh$. When $|k| = (3/2) k_{cross}$, advective AMF and gravitational AMF have equal contribution, and we denote that $|k|$ as $k_{eq}$. $k_{eq} = (3/2) k_{cross}  = 3/Qh$. When $|k|$ < $k_{eq}$, $F_G$ has larger contribution, whereas when $|k| > k_{eq}$, $F_A$ contributes more. Hence, to compare the relative importance of $F_A$ and $F_G$, we just need to compare the characteristic values of $|k|$ and $k_{eq}$.

Plugging Equations \ref{eq:toomreQ} and \ref{eq:kcrit} into the dispersion relation, 
\begin{equation}
    \Tilde{\omega}^2 = \Omega^2 - 2\pi G\Sigma|k| + c_s^2k^2,
    \label{eq:dispersion}
\end{equation}
where $\Tilde{\omega}(r) = m[\Omega_p - \Omega(r)]$ and $\Omega(r)$ is the angular velocity of the disk, we get the absolute value of the wave number in unit of $k_{crit}$,
\begin{equation}
    \frac{|k|}{k_{crit}} = \frac{2}{Q} \Bigg\{ m^2\Bigg(1-\frac{r^{3/2}}{r_p^{3/2}}\Bigg)^2 - \Big(1-\frac{1}{Q^2}\Big) \Bigg\}^{1/2} + \frac{2}{Q^2}.
    \label{eq:kmSG}
\end{equation}
Here we choose the positive sign in front of the first term since it corresponds to the short wave that propagates beyond Lindblad resonances in disks (the negative sign solution corresponds to the long wave that is restrained inside Lindblad resonances and outside the forbidden region; see Figure 2 in \citealt{lovelace97}, Figure 6.14 in \citealt{binney08} and Equation 19 in \citealt{goldreich79} for details). Note that this equation reduces to the following \citep{ogilvie02, bae2018a} as $Q \rightarrow \infty$,
\begin{equation}
    \frac{|k|}{m} = \frac{\Omega}{c_s} \Bigg|\Big(1 - \frac{r^{3/2}}{r_p^{3/2}}\Big)^2 - \frac{1}{m^2}\Bigg|^{1/2}.
    \label{eq:kmNOSG}
\end{equation}
With Equation \ref{eq:kmSG}, we evaluate the ratio of the $|k|$ and $k_{eq}$ at the effective locations of Lindblad resonances. While both sound pressure and self-gravity shift the positions of the Lindblad resonances \citep{pierens05}, we only account for the contribution from the pressure effect for simplicity. The locations of the resonances become, 
\begin{equation}
    r = r_p \Bigg\{ 1 \pm \frac{\big[1+m^2(h/r)^2\big]^{1/2}}{m} \Bigg\}^{2/3}.
\end{equation}
At these radii, the ratio reads,
\begin{equation}
    \frac{|k|}{k_{eq}} = \frac{Q}{3}\Bigg\{m^2(h/r)^2+\frac{1}{Q^2}\Bigg\}^{1/2} + \frac{1}{3}.
\end{equation}
As $Q \rightarrow \infty$, the ratio $|k|/k_{eq}$ $\rightarrow \infty$, which means that $F_G$ has no contribution to the total AMF. However, when $Q = 5$ and we adopt $m = (1/2)/(h/r)$, $|k| \approx k_{eq}$. Hence, $F_G$ contributes comparable amount of AMF as $F_A$. This simple calculation shows that as $Q$ becomes smaller, the gravitational AMF becomes increasingly important and will exceed the contribution of the advective  AMF at very low $Q$. 

We caution that the detailed analysis on the relative importance of $F_{A}$ and $F_{G}$ requires numerically solving the linearized equations and is beyond the scope of this paper. In this paper, since we focus on disks with $Q\gtrsim 5$, we will only discuss $F_A$ for simplicity. As shown in Section \ref{sec:unstable}, we find that when $Q\sim$2, the gap edge quickly becomes unstable even though the disk itself is gravitationally stable. 

\subsubsection{Calculation of the total AMF}
Here we summarize the calculation of the total AMF in the linear theory. 
In regions far from corotation ($|r-r_p|/r_p \rightarrow \infty$), advective AMF dominates ($F_A \gg F_G$, \citealt{goldreich80}). Since $F_J$ is conserved at different radii, $F_J$ equals $F_A$ far from corotation where WKB solutions (of Equation 83 therein) are valid. The analytical solution can be obtained \citep{goldreich78} at the following condition \citep{goldreich80},
\begin{equation}
    m (h/r) \ll \mathrm{min}\Big(\frac{Q^2-1}{3Q^2}, 1\Big).
    \label{eq:analyticalcon}
\end{equation}
The resulting AMF is given by,
\begin{equation}
        F_J^{WKB}(m) = \frac{4}{3}\frac{m^2\Sigma}{\Omega^2}\Big(\frac{GM_p}{R}\Big)^2 \{2K_0(2/3)+K_1(2/3)\}^2,
    \label{eq:AMFWKB}
\end{equation}
where the last term is around 6.35, and $K_0$ and $K_1$ are zeroth and first order modified Bessel function of the second kind. The value of $F_J^{WKB}(m)$ is approximately equal to 8.47 $m^2 F_{J0}$. However, the condition (\ref{eq:analyticalcon}) is not always satisfied in the disk. The analytical solution deviates from the exact solution when $m \gtrsim 1/(h/r)$, or disk self-gravity is important ($m \gtrsim \frac{Q^2-1}{3Q^2}$). At this regime,  \citet{goldreich80} numerically integrate the WKB trial solutions for each $m$ harmonic at given $Q$ and obtain $F_J(m, Q)$, expressed in unit of $F_J^{WKB}(m)$ (see Figures 2 and 3 therein and Figures \ref{fig:GT80Fig2Mp0p01} and \ref{fig:GT80Fig2Iso} in this paper). Summing up each $m$ mode, the total AMF is given by,
\begin{equation}
    F_{J} =  f(Q) \frac{F_J^{WKB}(m)}{m^2} \approx 8.47 f(Q) F_{J0},\label{eq:fQGT80}
\end{equation}
where $F_{J0}$ is expressed in Equation \ref{eq:FJ0}, and $f(Q)$ is a factor, which is a function of $Q$. From  \citet{goldreich80},
\begin{equation}
    f(Q) = \frac{1}{3} \mu^3_{max} = \int_{0}^{\infty}d\mu\mu^2\frac{F_J(m, Q)}{F_J^{WKB}(m)},
    \label{eq:Fratio}
\end{equation}
where $\mu = m c_s/\Omega r = m (h/r)$. The coefficient $F_J / F_{J0} \approx 0.93$ if $Q = \infty$ and $ F_J / F_{J0} \approx 8.6$ if $Q = 2$, which means that more massive disks (with lower Toomre $Q$ values) have higher AMF. We will compare this analytical solution with our simulation results in \ref{sec:resultlineartheory}.


\subsection{Pitch angle}
The pitch angle $\beta$ is defined as the angle between the direction of the line tangent to the spiral arm and the azimuthal direction in the disk. If the spiral arms have m-fold rotational symmetry, and $k$ is the radial wave number under the WKB approximation,
\begin{equation}
    \beta = tan^{-1}\big(\frac{m}{kr}\big).
\end{equation}
The dispersion relation for the spiral wave in the tightly wound limit (Equation \ref{eq:dispersion}) can also be used to derive the pitch angle.
In the absence of disk self-gravity, the middle term on the right hand side of Equation \ref{eq:dispersion} becomes zero. Thus, by manipulating Equation \ref{eq:kmNOSG}, the pitch angle for the $m$ mode is,
\begin{equation}
    tan(\beta) = \frac{m}{kr} = \frac{c_s}{\Big\{|\Omega_p - \Omega|^2 - \frac{\Omega^2}{m^2}\Big\}^{1/2}}.
    \label{eq:pitchanglenosg}
\end{equation}
If $m \gg 1$, different $m$ modes have the same pitch angle and interfere with each other \citep{ogilvie02}, so that
\begin{equation}
    tan(\beta) = \frac{m}{kr} = \frac{1}{r} \frac{c_s}{|\Omega_p - \Omega|}.
    \label{eq:pitchanglecommon}
\end{equation}
When the disk self-gravity is non-negligible, we can rewrite Equation \ref{eq:kmSG} to derive the pitch angle 
\begin{equation}
    tan(\beta) = \frac{1}{r} \frac{c_s}{\Big\{|\Omega_p - \Omega|^2 - \frac{\Omega^2}{m^2}\big(1-\frac{1}{Q^2}\big)\Big\}^{1/2} + \frac{\Omega}{mQ}}\,.
    \label{eq:pitchanglesg}
\end{equation}
Compared to Equation \ref{eq:pitchanglenosg}, two additional terms are $(\Omega/mQ)^2$ within the square root and $\Omega/mQ$ outside the square root. When $Q \gg 1$, Equation \ref{eq:pitchanglesg} reduces to Equation \ref{eq:pitchanglenosg}. Equation \ref{eq:pitchanglesg} also indicates that the pitch angle becomes smaller for higher disk masses (with smaller $Q$). However, this effect is not strong unless $Q$ is very close to unity. For example, when $m \gg 1$,  Equation \ref{eq:pitchanglesg} reduces to Equation \ref{eq:pitchanglecommon} even in massive disks. Thus, different $m$ modes should still have the same pitch angle and interfere with each other unless $Q$ is close to one. The effect should be stronger in the inner disk than the outer disk since $\Omega$ decreases with $r$. We will confirm this pitch angle calculation using our simulations in Section \ref{sec:resultSG}.

\subsection{Orbital cooling}

 \citet{miranda19b} carry out linear analysis on locally isothermal disks with temperature varying with the radius and run simulations to show that the spiral wave absorbs AMF from the background when it is propagating to hotter regions, resulting in higher AMF and higher density perturbation. The phenomenon is unique to locally isothermal disks. It would not happen even if the adiabatic disk is very close to locally isothermal ($\gamma$ is very close to 1). They advocate that the inclusion of the energy equation would result in weaker density waves and shallower gaps. Thus, planet masses inferred from previous locally isothermal simulations might be systematically lower than their actual masses.

Besides running locally isothermal simulations, we further study this issue by carrying out adiabatic simulations with a simple orbital cooling prescription. 
The adiabatic EoS is 
\begin{equation}
    P = (\gamma - 1)E.
\end{equation}
where $E$ is the internal energy per unit area and $\gamma$ is the adiabatic index. We adopt $\gamma$ = 1.4 in this study. Since the energy equation is solved in simulations with the adiabatic EoS, we can prescribe the effect of radiative cooling using the orbital cooling approach:
\begin{equation}
    \frac{dE}{dt} = -\frac{E - c_v \Sigma T_{irr}}{t_{cool}},
    \label{eq:cooling}
\end{equation}
where $\Sigma$ is the disk surface density. $T_{irr}$ is the disk initial temperature, which is determined by the stellar irradiation in realistic disks. $c_v \equiv R/(\mu (\gamma -1))$ is the heat capacity per unit mass, $R\equiv k/m_H$ is the specific gas constant, $k$ is the Boltzmann constant, $\mu$ is the mean molecular weight, $m_H$ is the atomic unit mass, and $t_{cool}$ is the cooling time. It is useful to define the dimensionless cooling time 
\begin{equation}
    T_{cool} = t_{cool}\Omega(r).
\end{equation}
Thus, $T_{cool}$ is $t_{cool}$ in the unit of orbital time over $2\pi$. This prescription is identical to \citet{zhu2015} and similar to the $\beta$-cooling in  \citet{gammie01}. Small $T_{cool}$ means fast cooling. Thus, the adiabatic disk with fast cooling should be closer to the isothermal disk, whereas large $T_{cool}$ means slow cooling, and this disk is closer to the adiabatic disk without cooling. In the rest of the paper, we simply use adiabatic disks to refer to the adiabatic disks without cooling.

To estimate $T_{cool}$ in realistic disks, we follow  \citet{zhu2015} which use the radiative cooling rate of
\begin{equation}
    \frac{dE}{dt} = -\frac{16}{3}\sigma(T_{mid}^4 - T_{irr}^4)\frac{\tau}{1+\tau^2},
    \label{eq:grayatm}
\end{equation}
where $\sigma$ is the Stefan-Boltzmann constant, $\tau = (\Sigma/2)\kappa_R$ is the optical depth in the vertical direction, $\kappa_R$ is the Rosseland mean opacity normalized to the gas surface density assuming dust-to-gas ratio is 1/100, $\kappa_R = \kappa_R(a_{max}, T)$, and $T_{mid}$ is the midplane temperature. Assuming $E = c_v\Sigma T_{mid}$ and using Equations \ref{eq:cooling} and \ref{eq:grayatm}, we can derive
\begin{equation}
    t_{cool} = \frac{3\Sigma c_v}{16\sigma (T_{mid}^2+T^2_{irr})(T_{mid}+T_{irr})}\frac{1+\tau^2}{\tau}.
    \label{eq:tcools}
\end{equation}

Approximating $T_{mid} = T_{irr} = T_d$, where
\begin{equation}
    T_d(r) = \begin{cases} 
      \Big(\frac{\phi L_*}{8\pi r^2 \sigma} \Big)^{1/4} & r \leq r_{T_f} \\
      T_{f} & r > r_{T_{f}} \\
   \end{cases}
   \label{eq:disktemp}
\end{equation}
where $\phi$ = 0.02, representing flaring angle \citep{dullemond18b}, and $L_*$ is the stellar luminosity. We assume the disk temperature decreases as $r$ and at a given radius it reaches a minimum floor temperature $T_{f}$, setting by the background heating processes (e.g., cosmic rays). This radius is
\begin{equation}
    r_{T_f} = 39\ \mathrm{au}\ \Big(\frac{T_f}{20\ \mathrm{K}}\Big)^{-2} \Big(\frac{L_*}{L_\odot}\Big)^{1/2} \Big(\frac{\phi}{0.02}\Big)^{1/2}.
\end{equation}
The dimensionless cooling time becomes,
\begin{equation}
    \begin{split}
    & T_{cool} = 0.015 \Big(\frac{f}{0.01}\Big)^{-1} \Big(\frac{\kappa_R(a_{max}, T_d)}{1\ \mathrm{cm^2 g^{-1}}}\Big)^{-1} \Big(\frac{L_*}{L_\odot}\Big)^{-3/4} \Big(\frac{\phi}{0.02}\Big)^{-3/4}  \\
    & \times \Big(\frac{M_*}{M_\odot}\Big)^{1/2} (1+\tau^2) \begin{cases} 
      1 & r \leq r_{T_f} \\
      \big(\frac{r}{r_{T_f}}\big)^{-3/2} & r > r_{T_f}
   \end{cases},
    \end{split}
    \label{eq:Tcool}
\end{equation}
where $f$ is the dust-to-gas mass ratio. For a given protoplanetary disk interior to $r_{T_f}$, and assuming constant $f$, $T_{cool}$ only depends on the Rosseland mean opacity in the optically thin limit, whereas $T_{cool} \propto \tau^2/\kappa_R$ in the optically thick limit. Within $r_{T_f}$, it does not explicitly depend on $r$  (the Rosseland mean opacity depends on the temperature which varies with the radius). Note that this optical depth is different from the optical depth in the dust continuum observation, since the Rosseland mean opacity is used for the cooling process. Aside from the optical depth effect, higher opacity, higher stellar radiation, larger flaring angle and smaller stellar mass would make $T_{cool}$ smaller and vice versa. We also ignore energy diffusion in the radial direction \citep{goodman2001}, which might be important for very optically thick disks.


Here we provide several typical values of $\kappa_R$ at different temperatures for references. Using the DSHARP opacity \citep{birnstiel18}, the Rosseland mean absorption opacity for the dust population with the maximum dust size $a_{max}$ = 1 $\mathrm{mm}$ and the size distribution $n(a) \propto a^{-3.5}$ is $\kappa_R(20\ \mathrm{K}) = 0.21\ \mathrm{cm^2 g^{-1}}$, $\kappa_R(40\ \mathrm{K}) = 0.48\ \mathrm{cm^2 g^{-1}}$ and $\kappa_R(125\ \mathrm{K}) = 1.03\ \mathrm{cm^2 g^{-1}}$.

For a solar-type star, suppose $\Sigma = 100\ \mathrm{g\ cm^{-2}}$, $f = 0.01$ at 10 au and $T = 40$ K, $T_{cool} \sim 20$, which is approaching the adiabatic limit. Suppose $\Sigma = 1\ \mathrm{g\ cm^{-2}}$, $f = 0.01$ at 100 au, and $T = 20$ K, $T_{cool} \sim 0.02$, which is on the locally isothermal limit. Thus, the $T_{cool}$ estimated from the gap substructures can help constrain the optical depth thus the surface density of the disk. We will use this argument to constrain the surface density at $\sim 100$ au of the AS 209 disk in Section \ref{sec:AS209}.

\section{Method} \label{sec:method}
To test the effects of disk self-gravity and radiative cooling on planet-disk interactions, we run four sets of 2D hydrodynamical simulations. We use FARGO-ADSG \citep{baruteau2008a, baruteau2008b, baruteau2016} in the first three sets and Athena++ (Stone et al. 2020, in prep) in the fourth set. For all the simulations, we fix the planet in circular orbits and do not consider planet migration. We apply the evanescent boundary condition to all the simulations. We choose not to include the indirect term for separating its effect from the effects studied in this paper. Our simulation setups are motivated and based on \citet{bae2018a} and  \citet{bae2017}. We keep all the parameters the same but add disk self-gravity or/and radiative cooling. We denote them as \texttt{B18} (Set 1) and \texttt{B17} (Set 2), respectively. We denote simulations with adiabatic EoS and radiative cooling as \texttt{AD}, and simulations with self-gravity as \texttt{SG}. For example, a simulation that is built on \citet{bae2017} with both orbital cooling and self-gravity is written as \texttt{B17ADSG}. We also run a set of simulations \texttt{AS209} (Set 3) including radiative cooling and self-gravity with massive planets by adopting the parameters in the AS 209 $\alpha$ varying model in \citet{zhang18} (\{$(h/r)_p$, $\alpha$, $M_p$\} = \{0.05, $3\times 10^{-4} (r/r_p)^2$, 1 $M_{th}$\}). The \texttt{Athena++} (Stone et al., 2020 in prep) simulations (Set 4)  have the same setup as \texttt{B17} and \texttt{B17AD} to verify the results. All the simulations that appear in this paper are summarized in Table \ref{tab:models}, and descriptions are detailed as follows. 

\begin{table*}
	\centering
	\caption{Models.}
	\begin{threeparttable}
	\label{tab:models}
	\begin{tabular}{lccccccccc} 
		\hline
		\hline
		FARGO-ADSG \\
		\hline
		Run & $Q$ & $T_{cool}$ & $M_p/M_{th}$ & $\alpha$ & $(h/r)_p$ & $q$  & Domain ($R$) & Resolution ($R \times \phi$) & $t_{grow}/t_p$\\
		\hline
		B18 & $\infty$ & 0 & 0.01 & 0 & 0.1 & 0.5  & [0.05, 5] & 4096 $\times$ 5580 & 1\\
		
		B18SG & 100 & 0 & 0.01 & 0 & 0.1 & 0.5  & [0.05, 5] & 4096 $\times$ 5580 & 1\\
		B18SG & 10 & 0 & 0.01 & 0 & 0.1 & 0.5  & [0.05, 5] & 4096 $\times$ 5580 & 1\\
		B18SG & 5 & 0 & 0.01 & 0 & 0.1 & 0.5  & [0.05, 5] & 4096 $\times$ 5580 & 1\\
		B18SG & 2 & 0 & 0.01 & 0 & 0.1 & 0.5  & [0.05, 5] & 4096 $\times$ 5580 & 1\\
		
		B18AD & $\infty$ & $10^{-4}$ & 0.01 & 0 & 0.1 & 0.5  & [0.05, 5] & 2048 $\times$ 2790 & 1\\
		B18AD & $\infty$ & $10^{-3}$ & 0.01 & 0 & 0.1 & 0.5  & [0.05, 5] & 2048 $\times$ 2790 & 1\\
		B18AD & $\infty$ & 0.01 & 0.01 & 0 & 0.1 & 0.5  & [0.05, 5] & 2048 $\times$ 2790 & 1\\
		\hline
		B17 & $\infty$ & 0 & 0.1 &  $5 \times 10^{-5}$ & 0.07 & 0  & [0.2, 2] & 2048 $\times$ 5580  & 20\\
		B17SG & 100 & 0 & 0.1 &  $5 \times 10^{-5}$ & 0.07 & 0  & [0.2, 2] & 2048 $\times$ 5580  & 20\\
		B17SG & 10 & 0 & 0.1 &  $5 \times 10^{-5}$ & 0.07 & 0  & [0.2, 2] & 2048 $\times$ 5580  & 20\\
		B17SG & 5 & 0 & 0.1 &  $5 \times 10^{-5}$ & 0.07 & 0  & [0.2, 2] & 2048 $\times$ 5580  & 20\\
		B17SG & 2 & 0 & 0.1 &  $5 \times 10^{-5}$ & 0.07 & 0  & [0.2, 2] & 2048 $\times$ 5580  & 20\\
	    B17SG & 100 & 0 & 0.3 &  $5 \times 10^{-5}$ & 0.07 & 0  & [0.2, 2] & 2048 $\times$ 5580  & 10, 60\\
		B17SG & 10 & 0 & 0.3 &  $5 \times 10^{-5}$ & 0.07 & 0  & [0.2, 2] & 2048 $\times$ 5580  & 10, 60\\
		B17SG & 5 & 0 & 0.3 &  $5 \times 10^{-5}$ & 0.07 & 0  & [0.2, 2] & 2048 $\times$ 5580  & 10, 60\\
		B17SG & 2 & 0 & 0.3 &  $5 \times 10^{-5}$ & 0.07 & 0  & [0.2, 2] & 2048 $\times$ 5580  & 10, 60\\
		B17SG & 100 & 0 & 1 &  $5 \times 10^{-5}$ & 0.07 & 0  & [0.2, 2] & 2048 $\times$ 5580  & 100\\
		B17SG & 10 & 0 & 1 &  $5 \times 10^{-5}$ & 0.07 & 0  & [0.2, 2] & 2048 $\times$ 5580 & 100\\
		B17SG & 5 & 0 & 1 &  $5 \times 10^{-5}$ & 0.07 & 0  & [0.2, 2] & 2048 $\times$ 5580 & 100\\
		B17SG & 2 & 0 & 1 &  $5 \times 10^{-5}$ & 0.07 & 0  & [0.2, 2] & 2048 $\times$ 5580 & 100\\
		B17SG & 100 & 0 & 3 &  $5 \times 10^{-5}$ & 0.07 & 0  & [0.2, 2] & 2048 $\times$ 5580 & 100\\
		B17SG & 10 & 0 & 3 &  $5 \times 10^{-5}$ & 0.07 & 0  & [0.2, 2] & 2048 $\times$ 5580 & 100\\
		B17SG & 5 & 0 & 3 &  $5 \times 10^{-5}$ & 0.07 & 0  & [0.2, 2] & 2048 $\times$ 5580 & 100\\
		B17SG & 2 & 0 & 3 &  $5 \times 10^{-5}$ & 0.07 & 0  & [0.2, 2] & 2048 $\times$ 5580 & 100\\
		B17AD & $\infty$ & 0.01 & 0.1 &  $5 \times 10^{-5}$ & 0.07 & 0  & [0.2, 2] & 2048 $\times$ 5580 & 20\\
		B17AD & $\infty$ & 0.1 & 0.1 &  $5 \times 10^{-5}$ & 0.07 & 0  & [0.2, 2] & 2048 $\times$ 5580 & 20\\
		B17AD & $\infty$ & 1 & 0.1 &  $5 \times 10^{-5}$ & 0.07 & 0  & [0.2, 2] & 2048 $\times$ 5580 & 20\\
		B17AD & $\infty$ & 10 & 0.1 &  $5 \times 10^{-5}$ & 0.07 & 0  & [0.2, 2] & 2048 $\times$ 5580 & 20\\
		B17AD & $\infty$ & 100 & 0.1 &  $5 \times 10^{-5}$ & 0.07 & 0  & [0.2, 2] & 2048 $\times$ 5580 & 20\\
		B17ADSG & 5 & 0.01 & 0.1 &  $5 \times 10^{-5}$ & 0.07 & 0  & [0.2, 2] & 2048 $\times$ 5580 & 20\\
		B17ADSG & 5 & 0.1 & 0.1 &  $5 \times 10^{-5}$ & 0.07 & 0  & [0.2, 2] & 2048 $\times$ 5580 & 20\\
		B17ADSG & 5 & 1 & 0.1 &  $5 \times 10^{-5}$ & 0.07 & 0  & [0.2, 2] & 2048 $\times$ 5580 & 20\\
		B17ADSG & 5 & 10 & 0.1 &  $5 \times 10^{-5}$ & 0.07 & 0  & [0.2, 2] & 2048 $\times$ 5580 & 20\\
		B17ADSG & 5 & 100 & 0.1 &  $5 \times 10^{-5}$ & 0.07 & 0  & [0.2, 2] & 2048 $\times$ 5580 & 20\\
		\hline
		AS209AD & $\infty$ & 0.01 & 1 &  $3 \times 10^{-4} (r/r_p)^2$ & 0.05 & 0  & [0.2, 2] & 1024 $\times$ 2790 & 20\\
		AS209AD & $\infty$ & 0.1 & 1 &  $3 \times 10^{-4} (r/r_p)^2$ & 0.05 & 0  & [0.2, 2] & 1024 $\times$ 2790 & 20\\
		AS209AD & $\infty$ & 1 & 1 &  $3 \times 10^{-4} (r/r_p)^2$ & 0.05 & 0  & [0.2, 2] & 1024 $\times$ 2790 & 20\\
		AS209AD & $\infty$ & 10 & 1 &  $3 \times 10^{-4} (r/r_p)^2$ & 0.05 & 0  & [0.2, 2] & 1024 $\times$ 2790 & 20\\
		AS209AD & $\infty$ & 100 & 1 &  $3 \times 10^{-4} (r/r_p)^2$ & 0.05 & 0  & [0.2, 2] & 1024 $\times$ 2790 & 20\\
		
		AS209ADSG & 5 & 0.01 & 1 &  $3 \times 10^{-4} (r/r_p)^2$ & 0.05 & 0  & [0.2, 2] & 1024 $\times$ 2790 & 20\\
		AS209ADSG & 5 & 0.1 & 1 &  $3 \times 10^{-4} (r/r_p)^2$ & 0.05 & 0  & [0.2, 2] & 1024 $\times$ 2790 & 20\\
		AS209ADSG & 5 & 1 & 1 &  $3 \times 10^{-4} (r/r_p)^2$ & 0.05 & 0  & [0.2, 2] & 1024 $\times$ 2790 & 20\\
		AS209ADSG & 5 & 10 & 1 &  $3 \times 10^{-4} (r/r_p)^2$ & 0.05 & 0  & [0.2, 2] & 1024 $\times$ 2790 & 20\\
		AS209ADSG & 5 & 100 & 1 &  $3 \times 10^{-4} (r/r_p)^2$ & 0.05 & 0  & [0.2, 2] & 1024 $\times$ 2790 & 20\\
		\hline
		Athena++ \\
		\hline
		B17 & $\infty$ & 0 & 0.1 &  $5 \times 10^{-5}$ & 0.07 & 0  & [0.2, 2] & 1024 $\times$ 2796 & 5\\
		B17AD & $\infty$ & 0.01 & 0.1 &  $5 \times 10^{-5}$ & 0.07 & 0  & [0.2, 2] & 1024 $\times$ 2796 & 5\\
		B17AD & $\infty$ & 0.1 & 0.1 &  $5 \times 10^{-5}$ & 0.07 & 0  & [0.2, 2] & 1024 $\times$ 2796 & 5\\
		B17AD & $\infty$ & 1 & 0.1 &  $5 \times 10^{-5}$ & 0.07 & 0  & [0.2, 2] & 1024 $\times$ 2796 & 5\\
		B17AD & $\infty$ & 10 & 0.1 &  $5 \times 10^{-5}$ & 0.07 & 0  & [0.2, 2] & 1024 $\times$ 2796 & 5\\
		B17AD & $\infty$ & 100 & 0.1 &  $5 \times 10^{-5}$ & 0.07 & 0  & [0.2, 2] & 1024 $\times$ 2796 & 5\\
		\hline
	\end{tabular}
	    \begin{tablenotes}
	       \small
           \item The $q$ is the exponent in the temperature profile, $T(r) \propto r^{-q}$. In \texttt{AD} models, the $(h/r)_p$ and the temperature profile are just initial values.
        \end{tablenotes}
	\end{threeparttable}
\end{table*}

(1) \texttt{B18} focuses on the linear regime of planet-disk interactions with very low mass planets. This setup is
also adopted in \citet{miranda19b}. The planet mass $M_p$ = 0.01 $M_{th}$. The aspect ratio at $r_p$ is $(h/r)_p$ = 0.1. The gas surface density is $\Sigma \propto r^{-1}$. The disk temperature is $T(r) \propto r^{-0.5}$ ($h/r \propto r^{0.25}$) with the locally isothermal EoS. The disk viscosity $\alpha$ is 0 (inviscid). The inner boundary is 0.05 $r_p$, whereas the outer boundary is 5.0 $r_p$.  The damping regions are inside 0.06 $r_p$ and outside 4.6 $r_p$. We run five simulations with $Q =$ $\infty$ (non-selfgravitating), 100, 10, 5, and 2, where $Q$ is evaluated at $r_p$. The self-gravity smoothing length is 0.3 $h$. Given the density and temperature profiles, $Q$ is larger at the inner disk and smaller at the outer disk. $Q \propto r^{-4/3}$ and $Q$ at the inner (outer) boundary is 9.5 times (0.30 times) the value at $r_p$. The resolution in (logarithmically spaced) $r$ and $\phi$ directions is 4096 $\times$ 5580. There are $\sim$ 89 grid cells per scale height in the $r$ direction. The planet potential is in the form of Equation \ref{eq:planetpotential} and the smoothing length $s$ is 0.6 $h$. The implementation of the Poisson equation solver for self-gravity can be found in \citet{baruteau2008b}. To study the AMF in simulations with adiabatic EoS and temperature varying with the radius, we also run three simulations (\texttt{B18AD}) with lower resolution (2048 $\times$ 2790) for $T_{cool} = 10^{-4}$, $10^{-3}$ and 0.01 cases.

(2) \texttt{B17} focuses on gap opening by more massive planets. These simulations start with globally isothermal disks. While one subset keeps the temperature constant, the other subset (\texttt{AD}) allows the disk to cool. The aspect ratio $(h/r)_{p}$ = 0.07, with $h/r \propto r^{0.5}$ so that $T$ is a constant.  The disk viscosity $\alpha$ = 5 $\times$ $10^{-5}$, and the gas surface density $\Sigma \propto r^{-1}$. The inner boundary is 0.2 $r_p$ and the outer boundary is 2.0 $r_p$. The damping regions are inside 0.24 $r_p$ and outside 1.6 $r_p$. The planet smoothing length $s$ is 0.1 $h$. If the self-gravity is included, the self-gravity smoothing length is 0.3  $h$. $Q \propto r^{-1/2}$ and $Q$ at the inner (outer) boundary is 2.2 times (0.71 times) the value at $r_p$. The resolution is 2048 (logarithmically spaced) $\times$ 5580 in the $r$ and $\phi$ directions, so that $dr$:$d\phi$ $\approx$ 1:1 for every grid in the domain. There are $\sim$ 62 grid cells per scale height in the $r$ direction. The planet mass grows with time as $M_p$ = $M_{p,f}$ $\mathrm{sin^2}(\frac{\pi}{2}\frac{t}{t_{grow}})$, and  $t_{grow}$ is 20 $t_p$ for 0.1 $M_{th}$, 60 $t_p$ for 0.3 $M_{th}$ and 100 $t_p$ for 1 and 3 $M_{th}$, where $t_p = 2\pi/\Omega$ is the orbital period of the planet. The planet grows to its full mass at $t_{grow}$ and stays constant afterwards. We mainly use $M_p = 0.3\ M_{th}$ cases to study the gap opening, but some instability has developed for the $Q=2$ disk before its planet grows to the full mass, complicating the analysis. Thus, we add four simulations ($M_p = 0.3\ M_{th}$, and $Q$ = 100, 10, 5 and 2) with $t_{grow}$ = 10 $t_p$. This set of simulations with a  shortened planet growing time is almost identical to the previous one at the time step we choose to analyze, but now the planet grows to its full mass before the instability occurs. Overall, we have $Q =$ $\infty$, 100, 10, 5 and 2 disks, with $M_p$ = 0.1, 0.3, 1 and 3 $M_{th}$. 

This set of simulations also includes dust, represented by 200, 000 dust super particles of different sizes. The Stokes numbers ($St$) of the particles at $r_p$  ranges from 1.57 $\times 10^{-5}$ to 1.57. The setup for dust particles is identical to \citet{zhang18}. The Stokes number $St$ for particles (also called particles' dimensionless stopping time) is 
\begin{equation}
St=t_{stop}\Omega=\frac{\pi s \rho_{p}}{2 \Sigma_{gas}}=1.57\times10^{-3}\frac{\rho_{p}}{1 \mathrm{g \,cm^{-3}}}\frac{s}{1 \mathrm{mm}}\frac{100 \mathrm{g\, cm^{-2}}}{\Sigma_{g}}\,.\label{eq:stokes}
\end{equation}
where $\rho_p$ is the density of the dust particle, $s$ is the radius of the dust particle, 
and $\Sigma_{g}$ is the gas surface density. We assume $\rho_p$=1 g cm$^{-3}$ in our simulations. The Stokes number mentioned above is the Stokes number at the beginning of the simulations, $St_0$. As the disk evolves, the dust sizes are fixed, but as they can drift in the disk,  the Stokes number can change.

For \texttt{B17AD} and \texttt{B17ADSG} disks, $T_{cool}$ is constant everywhere in each simulation, but $t_{cool}$ varies as radius since $\Omega_k \propto r^{-3/2}$. We explore $T_{cool}$ = 0.01, 0.1, 1, 10 and 100, covering fast cooling to slow cooling. Note that for these simulations with cooling, the given $h/r$ and the temperature profiles above are just the initial values. Their values are subject to change as the simulations evolve. We also run a set of \texttt{B17ADSG} disks, which includes both radiative cooling and self-gravity. These disks have $Q = 5$ at the planet's position and $T_{cool}$ = 0.01, 0.1, 1, 10 and 100.

(3) \texttt{AS209} is used to test the effects of radiative cooling in the massive planet regime ($\sim$ thermal mass). We run a set of varying $\alpha$ ($\alpha$ = $3 \times 10^{-4}(r/r_p)^2$) models (\texttt{AS209AD}) with $T_{cool}$ = 0.01, 0.1, 1, 10 and 100. The  $(h/r)_{p}$ is 0.05 and $M_p$ is 1 $M_{th}$ or 0.1 $M_J$ if $M_* = M_\odot$. The resolution in the $r$ and $\phi$ directions is 1024 $\times$ 2790. With lower resolution, we are able to run the simulations longer. We also add a set of simulations that also include self-gravity (\texttt{AS209ADSG}). $Q$ is chosen to be 5 at $r_p$. The planet grows to its full mass at 20 $t_p$. Other parameters are the same as (2). 

(4) \texttt{Athena++} simulations are used to verify the results. They include an isothermal simulation and a suite of adiabatic simulations, with the same setup as $M_p$ = $0.1\ M_{th}$, \texttt{B17} and \texttt{B17AD} disks in (2), except for a shorter $t_{grow} = 5\ t_p$. To avoid a numerical artifact due to the small smoothing length, we use $s$ = 0.2 $h$ for $T_{cool} = 100$ (twice as large as what is used in the FARGO simulation).

\section{Results} \label{sec:results}
\subsection{Self-gravitating disks \label{sec:resultSG}}

\begin{figure*}
\includegraphics[width=\linewidth]{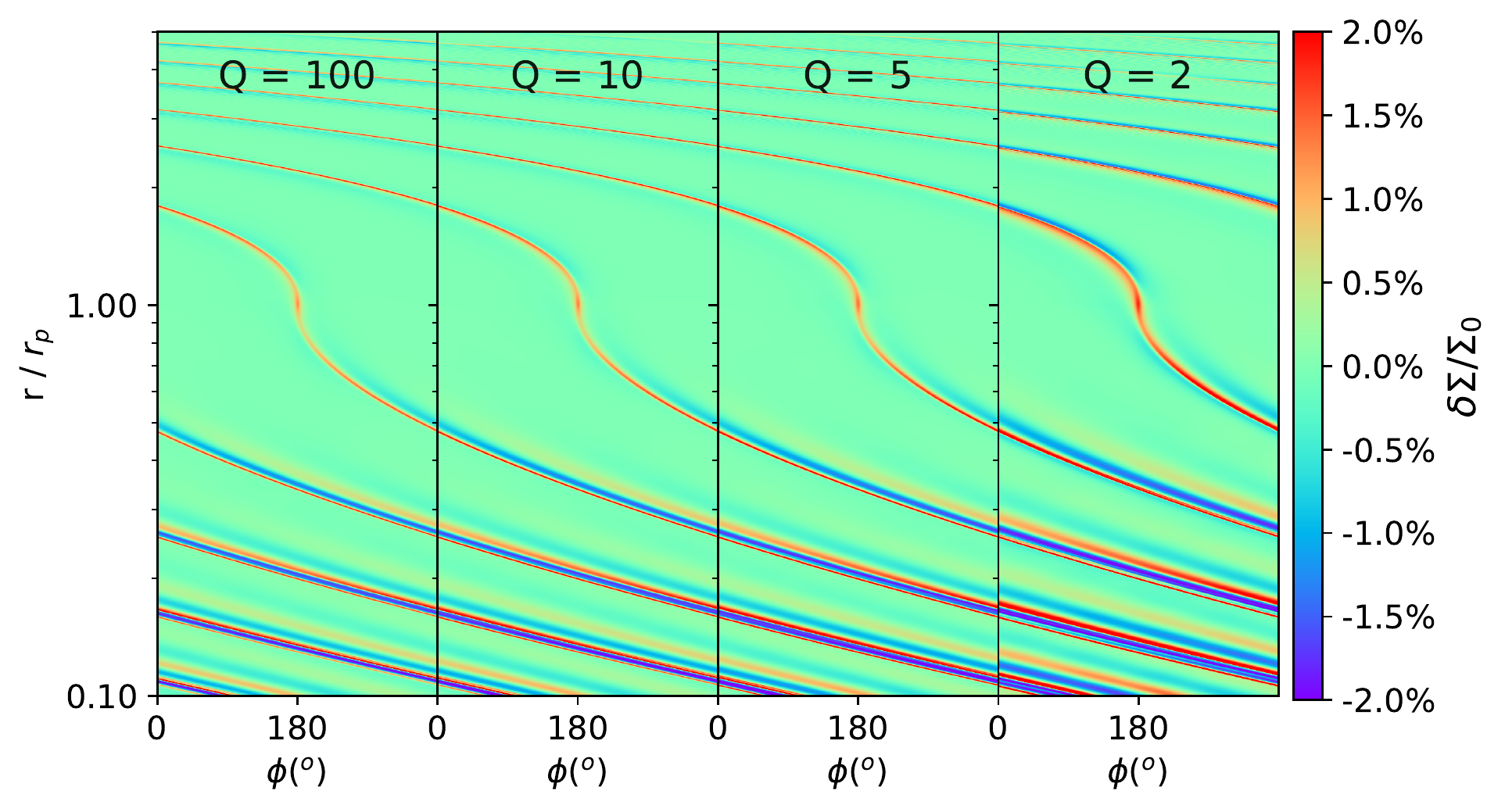}  
\caption{The gaseous density perturbations ($\delta\Sigma$/$\Sigma_0$) of the inviscid self-gravitating disks (\texttt{B18SG}) with a low mass planet $M_p$ = 0.01 $M_{th}$ at 10 orbits. The disk masses increase from left to right. At the position of the planet, the Toomre $Q$ parameters are 100, 10, 5 and 2, respectively. Since the density profile goes as $r^{-1}$, $Q$ is higher at the inner disk and lower at the outer disk. The spirals become stronger and tighter as the mass increases. This is the most evident in the $Q = 2$ disk.
\label{fig:BZ18_2Dgasdens}}
\end{figure*}

\begin{figure}
\includegraphics[width=\linewidth]{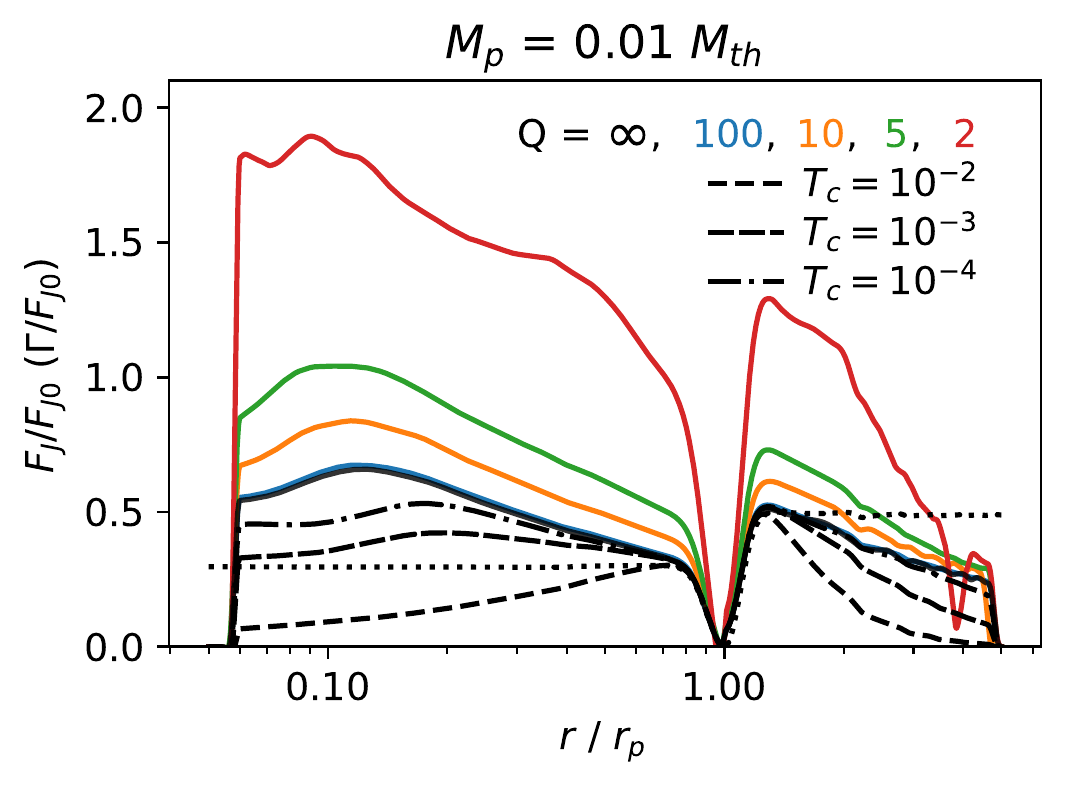}\caption{The Angular Momentum Flux (AMF) of the \texttt{B18}, \texttt{B18SG} and \texttt{B18AD} disks calculated from Equation \ref{eq:AMF} and normalized by Equation \ref{eq:FJ0}. The black curves represent the AMF (solid curve) and torque (dotted curve) of the non self-gravitating disk, which is almost identical to Figure 1(a) in \citet{miranda19b}. The AMF increases as the disk mass increases. The AMF of $Q=2$ disk is $\sim$ 3 times that of $Q=100$ disk. This results in a stronger density perturbation. Short dashed, long dashed, and dashed-dotted curves in black represent \texttt{B18AD} with $T_{cool}$ = 0.01, $10^{-3}$ and $10^{-4}$, and with lower resolution (2048 $\times$ 2790). The $T_{cool}$ = $10^{-4}$ curve is similar to the case without cooling at such resolution.
\label{fig:BZ18_1dDensAMF}}
\end{figure}

To study the effects of disk self-gravity in the linear regime, we first analyze \texttt{B18SG} disks with $M_p$ = 0.01 $M_{th}$. Figure \ref{fig:BZ18_2Dgasdens} shows  the density perturbations with different disk masses in the polar coordinate. The planet is at $\phi$ = 180$^\circ$ and $r$ = 1 $r_p$. The density perturbation is
\begin{equation}
    \delta\Sigma/\Sigma_0 = (\Sigma - \Sigma_0)/\Sigma_0,
\end{equation}
where $\Sigma_0$ is the gaseous surface density at the initial condition, and $\Sigma$ is the density at the time step (t = 10 $t_p$ for this plot) that is analyzed. From left to right, the disk mass increases. The Toomre $Q$ values at the $r = r_p$ are 100, 10, 5 and 2, respectively. As the disk mass increases, the spiral perturbations become stronger. This transition is the most evident from  the $Q=5$ to $Q=2$ disk. The density perturbation changes slightly until $Q=5$ and has a noticeable increase from the $Q=5$ to $Q=2$ disk. The underlying AMF that influences the density perturbation is shown Figure \ref{fig:BZ18_1dDensAMF}. The AMF increases as $Q$ decreases. The AMF at $Q = 100$ (blue solid curve) is almost identical to the one in the non-selfgravitating disk (black curve) and identical to the result in Figure 1(a) in  \citet{miranda19b}. As pointed out by \citet{miranda19b}, we can clearly see that the AMF increases towards the inner disk where the disk is hotter. However, it cannot be captured even with very small cooling time $T_{cool} = 0.01$ and 0.001, as shown by the short and long dashed curves. Different from the locally isothermal simulation, the spiral waves cannot pick up the AMF from the background disk when they propagate to hotter regions, even with a relatively short $T_{cool}$. Only extremely small cooling time ($T_{cool}$ = $10^{-4}$, shown by the dotted-dashed curve) can capture this increase of the AMF towards inner disk (the slightly lower values are due to the lower resolution). The jump of AMF is the largest from the $Q=5$ to $Q=2$ disk, where the AMF value of the latter almost doubles. The increase of the AMF with a smaller $Q$ is consistent with the fact that the density perturbation becomes stronger with a smaller $Q$.

The spirals become more tightly wound as the disk mass increases. It is the most evident at the inner disk between the $Q=5$ and $Q=2$ disk. This effect is consistent with the prediction in Section \ref{sec:theory} around Equation \ref{eq:pitchanglesg}. 
Only when $Q$ becomes very small, the two additional terms become important, which makes the pitch angle smaller. Note that $Q$ is larger than $Q_p$ in the inner disk, given the surface density profile we choose. The changes in pitch angle would be larger if $Q$ is constant across the disk. This trend is also reported in  \citet{pohl15}, where their simulations lie in the non-linear regime with higher planet masses. Hence, the decrease of pitch angle with the disk mass applies to a large range of planet masses. 

\citet{yu19} measure the pitch angles from 10 protoplanetary disks in near-infrared scattered-light images and three disks in ALMA millimeter dust continuum images. They find a tight relation between the spiral arm pitch angle and the disk mass---more massive disks have smaller pitch angles. While we find that the disk self-gravity cannot lead to such significant changes in pitch angles, spirals in scattered-light images might show more dramatic effects with disk self-gravity which needs to be studied in future. Without the consideration of the self-gravity, the different pitch angles between scattered light and millimeter continuum images might be due to the vertical temperature gradient \citep{juhasz18, rosotti19}. The near-infrared scattered-light observations probe higher atmosphere with higher temperature, thus the spirals have larger pitch angles than those in the midplane probed by ALMA.

\subsubsection{Comparison to the linear theory} \label{sec:resultlineartheory}
In this subsection, we provide a quantitative comparison with the linear theory. In Section \ref{sec:theory}, we briefly introduce the AMF in disks with and without self-gravity. The quantity we want to compare is $F_J(m, Q)/F_J^{WKB}(m)$ (ratio of the numerically integrated $m$ harmonics of the AMF and the $m$ mode AMF solved analytically; see Equation \ref{eq:Fratio}), which has been given in Figure 2 of \citet{goldreich80}. Similar comparisons have been performed before in non-selfgravitating shearing sheet simulations \citep{dong2011a}. With our simulations, we measure the AMF at several $h$ away from a low-mass planet ($M_p$ $\ll$ $M_{th}$) in the inner disk. Note that the spirals will shock at a certain distance away from the planet. This distance is $l_{sh}$ \citep{goodman2001}, given by
\begin{equation}
    l_{sh} \approx 0.8 \Big(\frac{\gamma+1}{12/5}\frac{M_p}{M_{th}}\Big)^{-2/5} h.
\end{equation} 
The linear theory of wave propagation fails for $|r-r_p|$ $>$ $l_{sh}$, so the position that we measure the AMF cannot be too far from the planet. The position should also be chosen closer to the planet as the planet mass increases, as $l_{sh}$ becomes shorter with higher mass planets.

\begin{figure} 
\includegraphics[width=\linewidth]{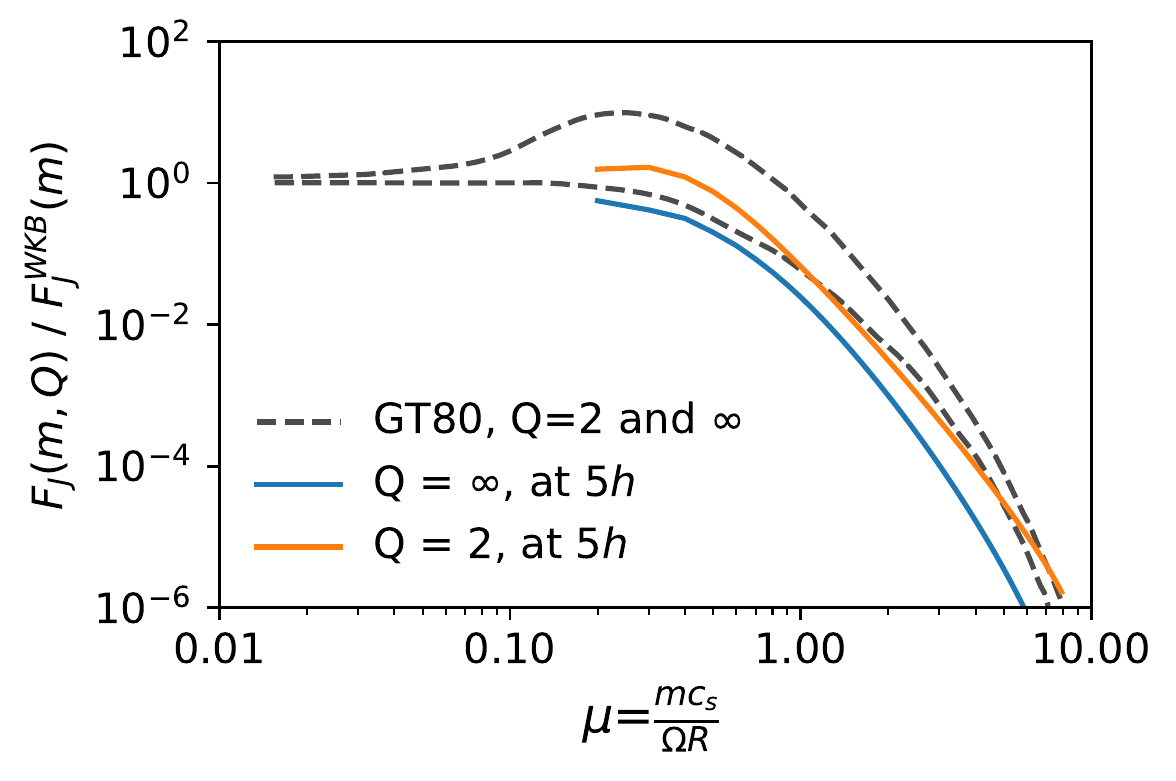} 
\caption{Comparison between $F_J(m, Q)/F_J^{WKB}(m)$ measured from simulations ($M_p$ = 0.01 $M_{th}$, \texttt{B18SG}) and calculated from the linear theory \citep{goldreich80}. This ratio is measured at $5 h$ (at 0.5 $r_p$) inside of the planet and is plotted on y-axis. The x-axis is $\mu = \frac{m c_s}{\Omega R}$ (or $m (h/r)$, where $(h/r)_p$ = 0.1). The simulation results are shown in blue ($Q=\infty$) and orange ($Q=2$) solid curves, whereas the analytical results are shown in dashed curves ($Q=2$ curve is the higher).
\label{fig:GT80Fig2Mp0p01}}
\end{figure}

We start the comparison with a very low planet mass, $M_p$ = 0.01 $M_{th}$ in \texttt{B18SG} disks. Figure \ref{fig:GT80Fig2Mp0p01} shows the AMF measured at 5$h$, where we find the values match the linear theory the best. The blue curve shows the ratio of the non-selfgravitating disk, and the orange curve shows the ratio of the $Q=2$ disk. The ratios calculated from  \citet{goldreich80} are shown in dashed curves (the ratio of $Q = 2$, $F_{J}(m, 2)/F_{J}^{WKB}(m)$ is shown above and that of $Q = \infty$, $F_{J}(m, \infty)/F_{J}^{WKB}(m)$ is shown below). The $Q = \infty$ curve matches the linear calculation relatively well, especially at $\mu$ $\lesssim$ 1. As for the differences, one of the reasons is that the temperature is assumed to be constant throughout the disk in the linear calculation, whereas $T\propto r^{-0.5}$ in the simulations. $F_J(m, Q)/F_J^{WKB}(m)$ of the self-gravitating disk is higher than that of $Q=\infty$ disk, which is consistent with the linear theory.

\begin{figure} 
\includegraphics[width=\linewidth]{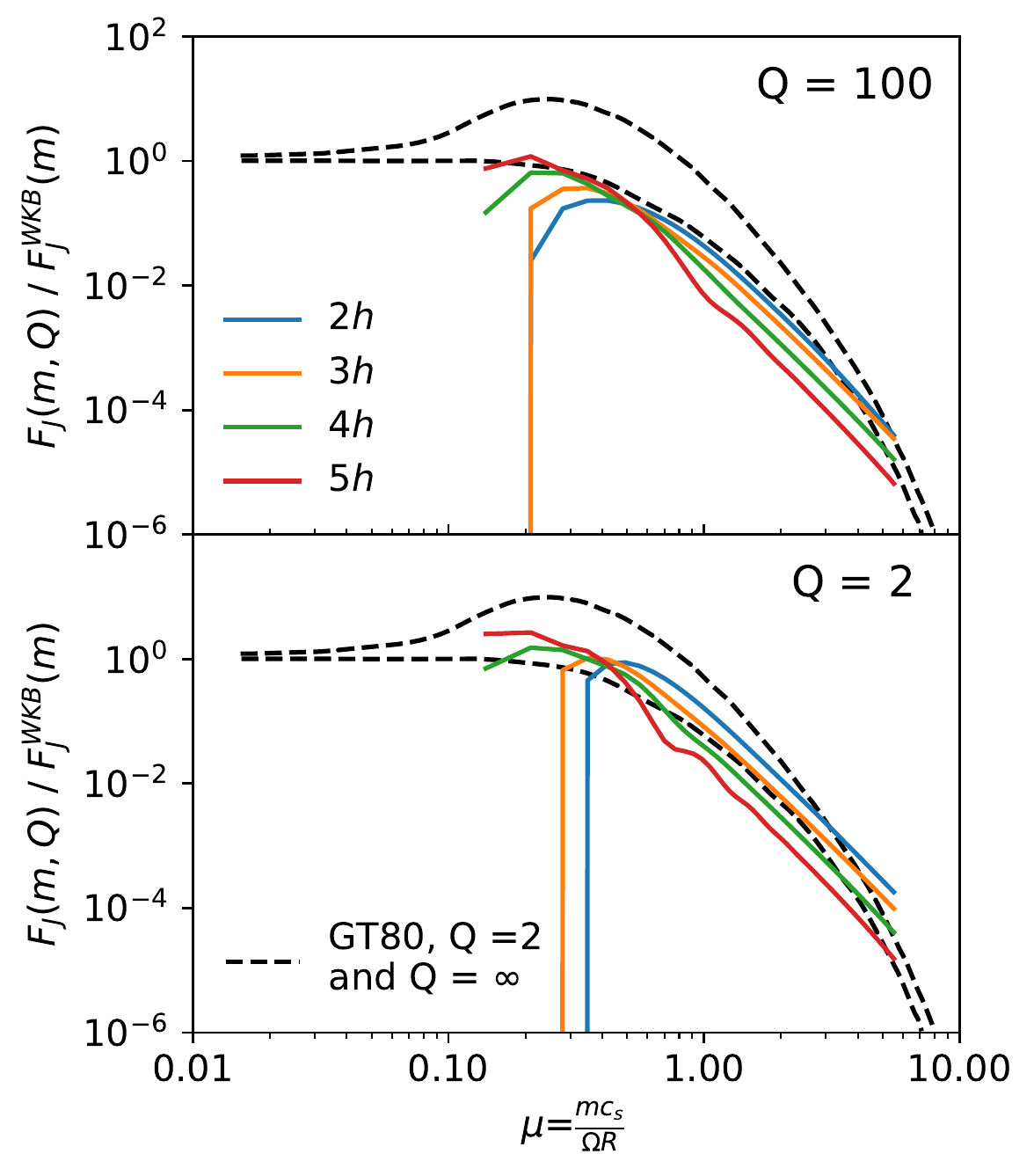}  
\caption{Similar to Figure \ref{fig:GT80Fig2Mp0p01}, but for \texttt{B17SG}, $M_p$ = 0.1 $M_{th}$. This ratio is measured at  $2 h$ (blue), $3 h$ (orange), $4 h$ (green), and $5 h$ (red) inside of the planet. The upper panel shows the results in $Q = 100$ disk, whereas the lower panel shows the results in $Q=2$ disk. The dashed curves are the same as in Figure \ref{fig:GT80Fig2Mp0p01}. 
\label{fig:GT80Fig2Iso}}
\end{figure}

To make a more proper comparison with the analytical theory, we also present \texttt{B17SG} disks that are globally isothermal with $M_p$ = 0.1 $M_{th}$. 
$F_{J}(m, Q)/F_{J}^{WKB}(m)$ at different distances from the planet are shown in Figure \ref{fig:GT80Fig2Iso} (Top panel: $Q=100$ disk,  Bottom panel: $Q=2$ disk). Similar to the results of isothermal shearing sheet simulations in \citet{dong2011a}, we also find a close agreement with the linear theory when $Q \rightarrow \infty$. However, the values in $Q = 2$ disk are still different from the linear theory expectation. They are lower by one order of magnitude when $\mu \lesssim 1$ and higher than the linear theory prediction at $\mu \sim 6$. The discrepancy in the high $m$ (or high $\mu$) is possibly due to the limited resolution in the azimuthal direction in simulations. We need very high resolutions to resolve the high $m$ modes properly with a much smaller self-gravity smoothing kernel. As for the low $m$ modes, it is actually not surprising to see discrepancy, since we only measure the advective AMF, and ignore the gravitational AMF (see Section \ref{sec:AMF}). In $Q = 2$ disks, $F_G$ should contribute a significant amount of AMF. It should be more important when $Q$ is even smaller (close to 1). To match the analytical result of total AMF, one has to include $F_G$. On the other hand, we do see the increases of advective AMF as the $Q$ becomes smaller. This indicates that as the disk becomes more massive, both $F_A$ and $F_G$ increase to contribute for the increase of total AMF, $F_J$. Being part of the AMF, $F_A$ alone is enough to explain the change of the density perturbations, i.e., the increases of $F_A$ leads to the increases of the density perturbations. It is not necessary to add $F_G$ when we want to explain the density perturbations qualitatively.

\subsubsection{Gap opening}
In the linear theory, if the planet mass $M_p$ is large enough (close to or higher than $M_{th}$), the planet can open gaps in the disk. However, the gap opening mass can be much lower, since the non-linear effects such as shock can occur even if the system starts in a linear regime \citep{goodman2001, rafikov02}. The study of gap opening is very important since gaps are observable in both gaseous and dusty disks. They have been found ubiquitously in recent high-resolution ALMA observations \citep{huang18b, long18}. Assuming these gaps are due to the planet, one might infer the planet mass from the gap width and depth (e.g., \citealt{zhang18}). Now we use higher mass planets to demonstrate the effect of the self-gravity on gap opening.

\begin{figure*} 
\includegraphics[width=\linewidth]{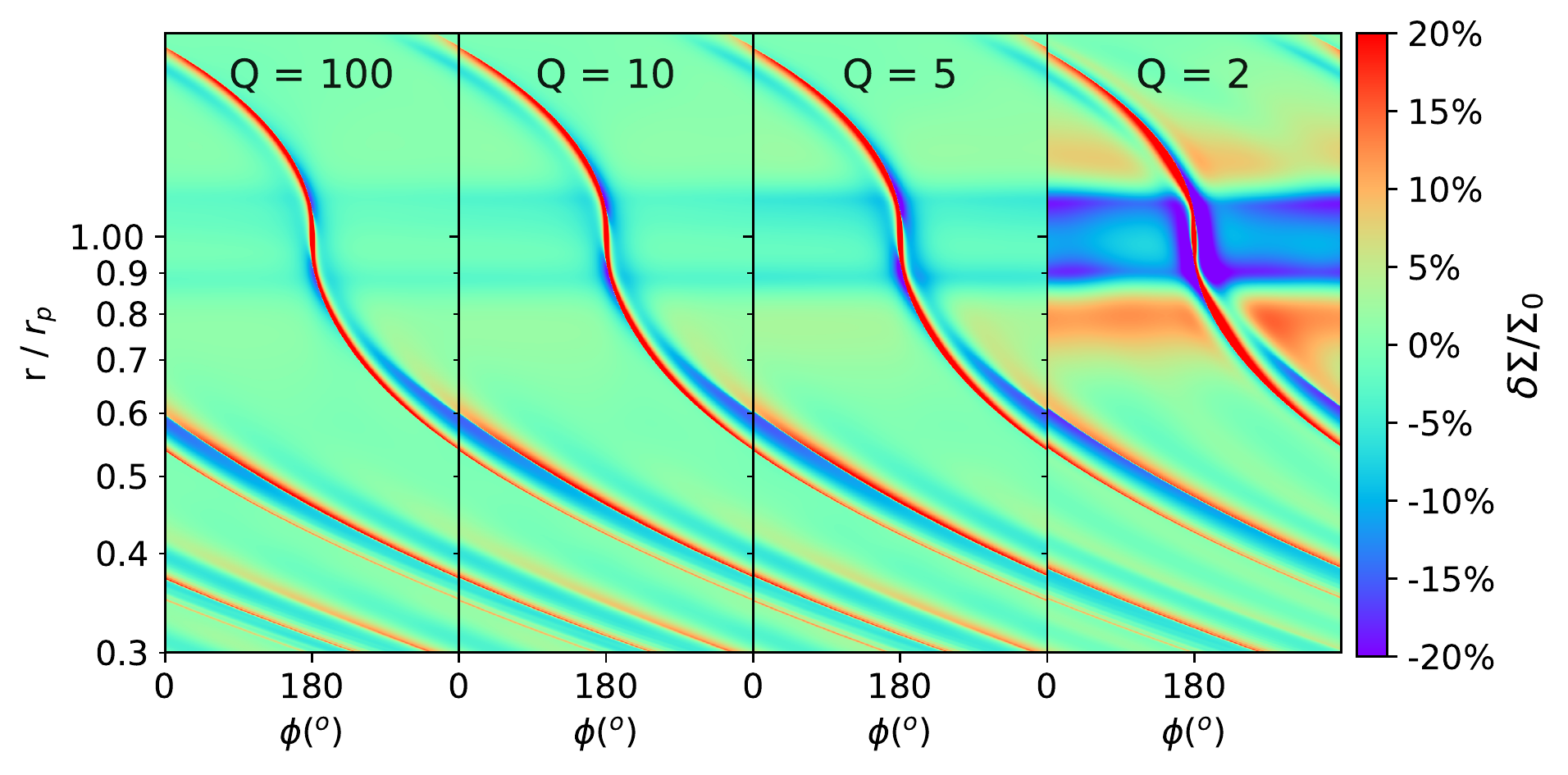}  
\caption{The gas density perturbations ($\delta\Sigma$/$\Sigma_0$) of the isothermal self-gravitating disks \texttt{B17SG} with a low mass planet $M_p$ = 0.3 $M_{th}$ at 20 orbits (with $t_{grow} = 10\ t_p$). Compared to Figure \ref{fig:BZ18_2Dgasdens}, higher planet mass can open deeper gap and even the secondary gap (see Figure \ref{fig:BZH17_1DgasdensMp0p3}). The depth of the primary gap becomes deeper with the decreases of $Q$. This is the most evident in $Q = 2$ disk.
\label{fig:BZH17_2DgasdensMp0p3}}
\end{figure*}

\begin{figure} 
\includegraphics[width=\linewidth]{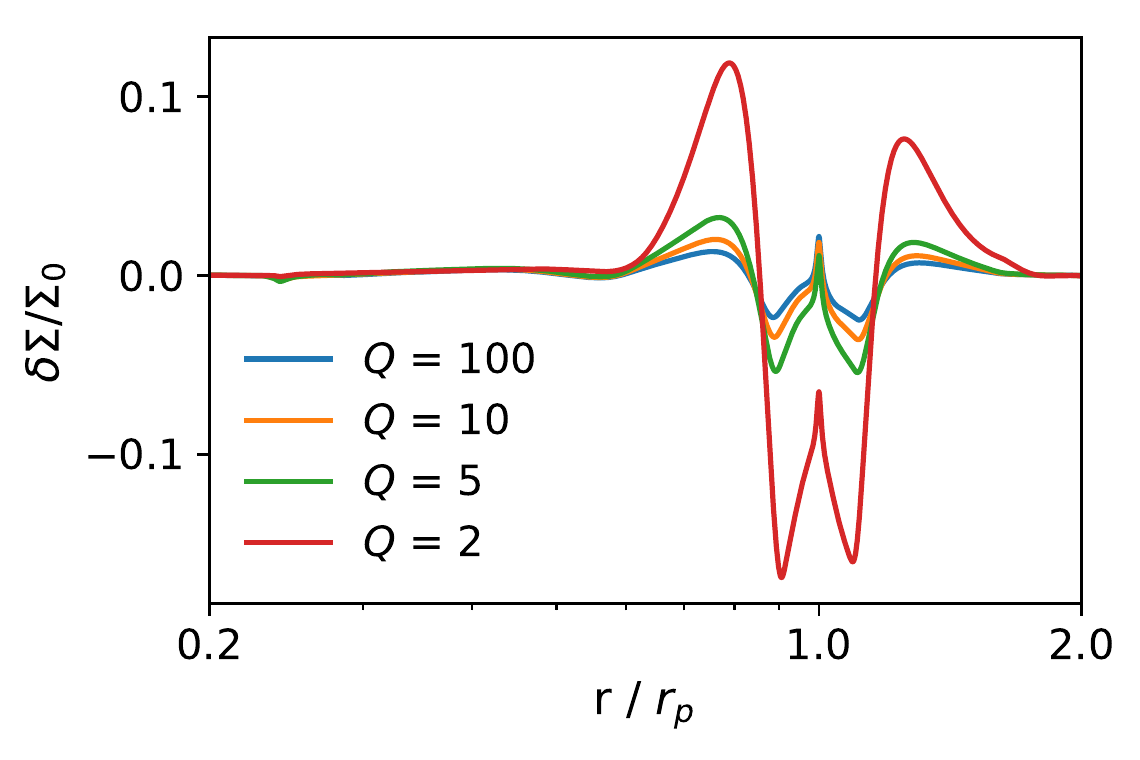}  
\caption{The azimutally averaged density perturbations of Figure \ref{fig:BZH17_2DgasdensMp0p3}. The blue, orange, green and red curves represent $Q =$ 100, 10, 5 and 2 disks. As the disk mass increases, the perturbations become stronger and gaps become deeper, but the position of the secondary gap does not change significantly.
\label{fig:BZH17_1DgasdensMp0p3}}
\end{figure}

We first show the results of \texttt{B17SG} disks with $M_p$ = 0.3 $M_{th}$, where the evolution starts in linear regime, marginally opening gap for non-selfgravitating disk at longer time evolution. Figures \ref{fig:BZH17_2DgasdensMp0p3} and \ref{fig:BZH17_1DgasdensMp0p3} show the density perturbations in 2D and 1D at $t = 20\ t_p$ (we use the models whose $t_{grow} = 10\ t_p$ here). As the $Q$ decreases, the gap becomes deeper. \citet{pohl15, li16} also find this effect for higher mass planets. The transition is the most evident from $Q=5$ to $Q=2$ disks, where the gap depth becomes a factor of 3 stronger at the major gap where the planet is located. The major gap is separated by the horseshoe region, thus resembles ``w'' shape. The gap width changes only slightly comparing to the gap depth (see Section \ref{sec:implications} for details). The secondary gap is around 0.5 to 0.6 $r_p$, the location of which does not change as the $Q$ decreases.

\begin{figure*} 
\includegraphics[width=\linewidth]{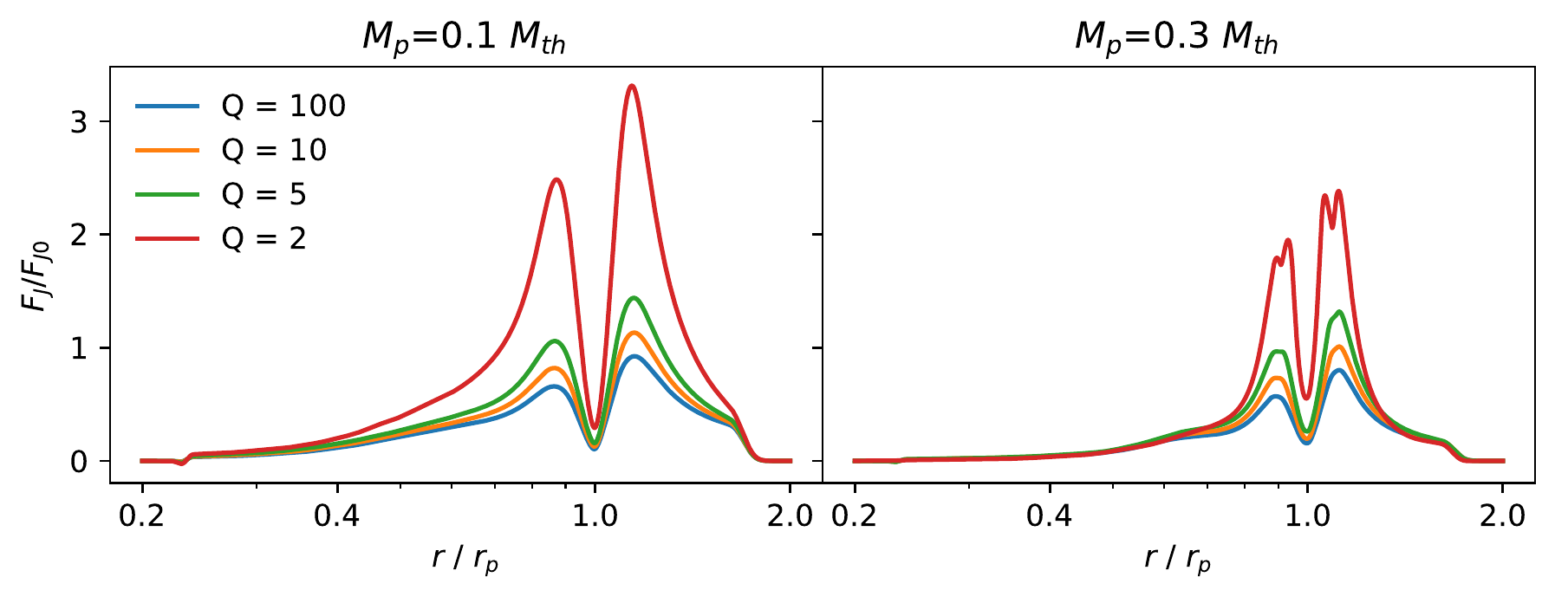}  
\caption{The AMF of the isothermal disks at 20 orbits with $M_p$ = 0.1 and 0.3 $M_{th}$. Given the same disk mass, as the planet mass increases, the normalized AMF becomes lower at any disk with different $Q$, as shock forms earlier and closer to the planet with the increases of the planet mass.
\label{fig:BZHAMF50orbits}}
\end{figure*}

To understand the density perturbations, we plot the AMF of each simulation in Figure \ref{fig:BZHAMF50orbits}. The left panel shows $M_p$ = 0.1 $M_{th}$ disks and right panel shows $M_p$ = 0.3 $M_{th}$ disks. Given the same $Q$, the normalized AMF decreases as the planet mass increases, which is due to the increasing non-linear shock dissipation. This is consistent with the result of non-selfgravitating disks in  \citet{miranda19b}. The changes of the AMF at each $Q$ has the same trend as the changes in density perturbations. From $Q=5$ to $Q=2$, the AMF doubles. We will quantify this trend in Section \ref{sec:AMFvs}. Since the normalization factor $F_{J0} \propto Mp^2$, the absolute value of AMF is still higher in the disks with higher planet mass.

\begin{figure*} 
\includegraphics[width=\linewidth]{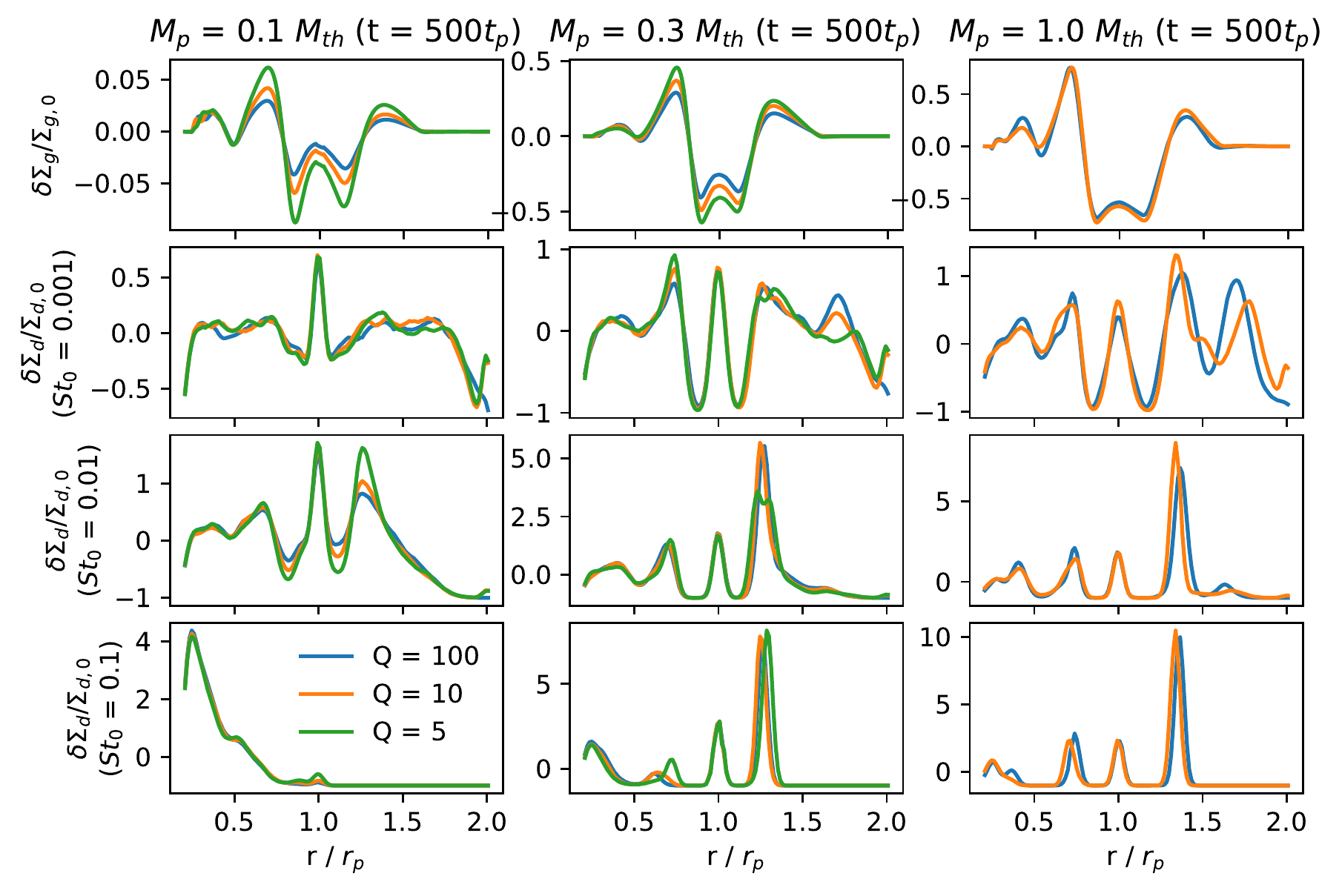}  
\caption{The density perturbations of the gas and dusts with different Stokes number ($St$ = 0.001, 0.01, and 0.1) at t = 500 $t_p$ in globally isothermal disk with $M_p$ = 0.1, 0.3 and 1 $M_{th}$ (from left to right). From top to bottom panels, there are density perturbations of the gas, $St= 0.001$, $St= 0.01$ and $St= 0.1$ sized dusts. Blue, orange and green curves represent $Q =$ 100, 10, and 5 disks. Some curves for disks with $Q=5$ are missing since the disk becomes unstable at this time steps in the simulations. Overall, the position of the secondary gap does not change with the increase of disk masses. The gap becomes deeper for lower $Q$ values.
\label{fig:BZHgasdustdens1d}}
\end{figure*}

To study how disk self-gravity affects the dust continuum observations, we evolve simulations for a longer time scale (500\ $t_p$) and plot the density perturbations of both gas and dust with different Stokes numbers in Figure \ref{fig:BZHgasdustdens1d}. From top to bottom, it shows the $\delta\Sigma/\Sigma_0$ of gas, and dust with $St$ = 0.001 (represented by $St$=[0.0005, 0.002]), 0.01 (represented by $St$=[0.005, 0.02]) and 0.1 (represented by $St$=[0.05, 0.2]). Notice that the $St_0$ is the Stokes number at the beginning of the simulations. As the simulations evolve, the particle sizes are fixed, so the dust grain might have different $St$ after it drifts. From left to right, the planet masses are $M_p$ = 0.1, 0.3 and 1.0 $M_{th}$. The blue, orange and green curves represent $Q$ = 100, 10, and 5 disks. As expected, the perturbations in the dust are stronger than those in gas, and dust with higher Stokes number has larger perturbations. This is because dust particles with higher $St$ drift faster when $St \lesssim 1$. For $M_p$ = 0.1 $M_{th}$, $St$ = 0.1 disk, the dust particles have almost drifted to the inner disk. Similar to the trend in the gas, the gap becomes deeper, and the ring becomes higher as the $Q$ becomes smaller, which are consistent with \citet{li16}. The positions of the secondary gaps do not change. The overall shapes of the profiles do not change, except the gaps outside the planet become weaker for $M_p$ = 0.3 and 1.0 $M_{th}$ disks at $St_0$ = 0.001. The positions of them also shift outside. The region beyond $r = 1.6\ r_p$ should not be trusted as it is in the wave damping zone. For small $Q$ and large $M_p$, some curves for disks with $Q=5$ are missing since the disks have already become unstable at $500\ t_p$. We will discuss this phenomenon in Section \ref{sec:unstable}.

\subsection{Adiabatic disks with radiative cooling }\label{sec:resultsAD}
\begin{figure*} 
\includegraphics[width=\linewidth]{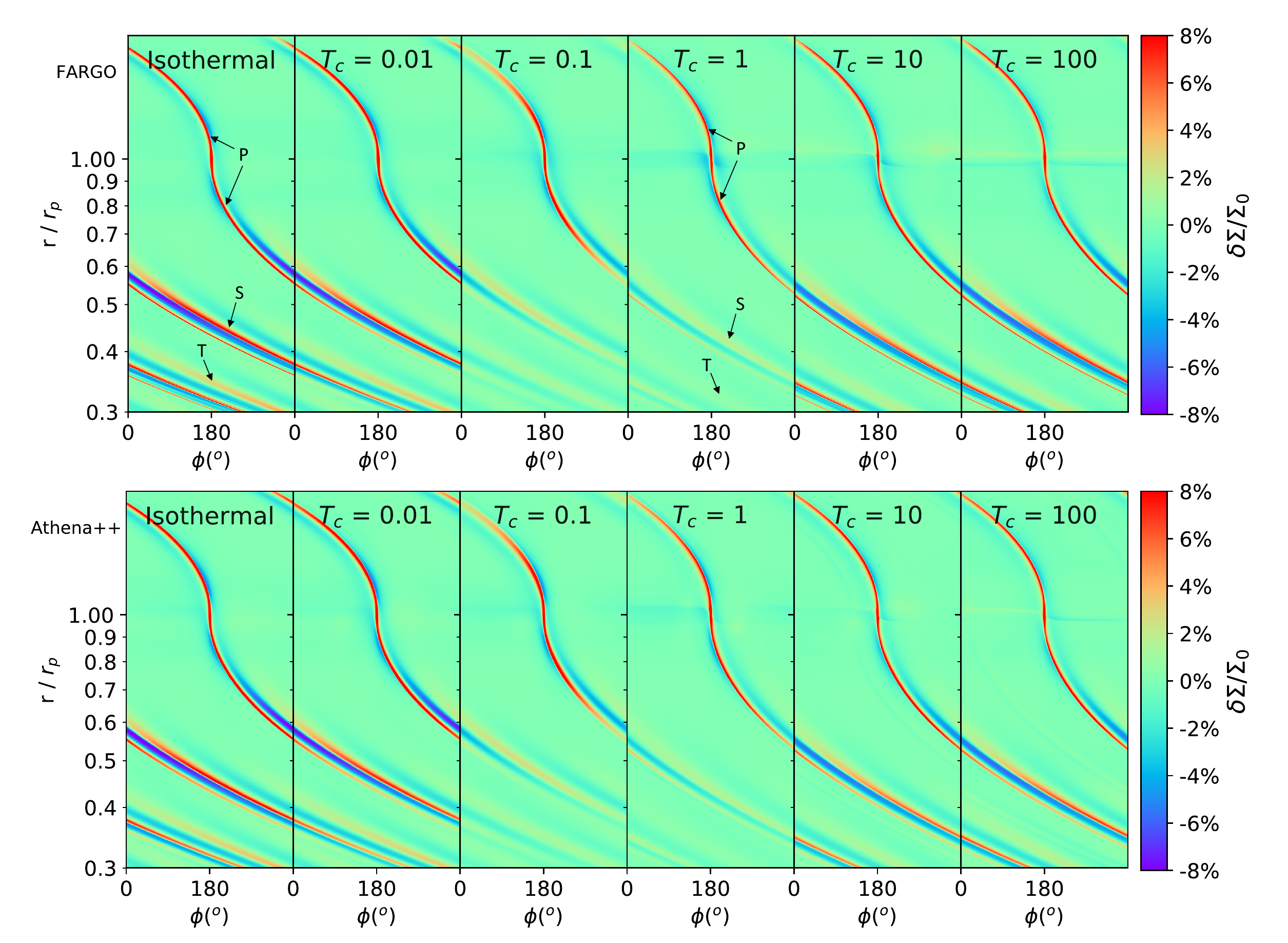}
\caption{The density perturbations of the adiabatic simulations \texttt{B17AD} at 40 orbits (FARGO) and 20 orbits (Athena++) with $M_p$ = 0.1 $M_{th}$. From left to right, the disks represent fast to slow cooling. The isothermal disk is shown in the leftmost panel, then the dimensionless disk cooling parameters $T_{cool}$ are 0.01, 0.1, 1, 10 and 100, respectively. The upper panels show results from FARGO-ADSG and the lower panels show results from Athena++. The primary spirals are marked as ``P'', the secondary spirals are marked as ``S'' and the tertiary spirals are marked as ``T''.
\label{fig:adiabatic2Dgasdens}}
\end{figure*}
In this subsection, we study the effects of radiative cooling on planet-disk interactions, focusing on \texttt{B17AD}. The simulations presented in this section are globally isothermal, with $q = 0$. The absorption of AMF from the background occurs in locally isothermal disk (in Figures \ref{fig:BZ18_2Dgasdens} and \ref{fig:BZ18_1dDensAMF}), but not in globally isothermal disks here. From left to right, Figure \ref{fig:adiabatic2Dgasdens} shows the density perturbations for the $M_p$ = 0.1 $M_{th}$ isothermal disk, and adiabatic disks with cooling time $T_{cool}$ = 0.01, 0.1, 1, 10 and 100. The top panels are results from FARGO-ADSG, whereas the bottom panels are results from Athena++. They are almost identical, except for the small difference for the corotation features in the $T_{cool} = 100$ cases. We analyze the simulations at $t$ = 40 $t_p$ for FARGO and 20 $t_p$ for Athena++ due to a shorter $t_{grow}$ adopted in Athena++ simulations. At these times, we find that all the disks are stable and have reached quasi-steady states. Disks with shorter dimensionless cooling times $T_{cool}$ are more similar to isothermal disks, whereas disks with larger $T_{cool}$ resemble adiabatic disks. The primary spirals are marked as ``P'', the secondary spirals are marked as ``S'' and the tertiary spirals are marked as ``T'' on the figure. 

From the isothermal to the adiabatic limit (increasing $T_{cool}$ from the left to the right panels), the spirals become weaker as the cooling timescale becomes longer until $T_{cool}$ = 1 and then become stronger again as $T_{cool}$ continues to increase. The secondary and tertiary spirals also become very weak at $T_{cool}$ = 1 and stronger at $T_{cool}$ = 10. Nevertheless, the secondary and tertiary spirals are still present at $T_{cool} = 1$ as indicated by the arrows. 

The spirals become more open as the cooling time increases. This is due to the increases of the sound speed from $c_s=(RT/\mu)^{1/2}$ in the isothermal limit to $c_s=(\gamma RT/\mu)^{1/2}$ in the adiabatic limit. In other words, the effective $\gamma$ in Equation \ref{eq:pitchanglecommon} increases from 1 to 1.4 as $T_{cool}$ becomes larger. 

\begin{figure*} 
\includegraphics[width=\linewidth]{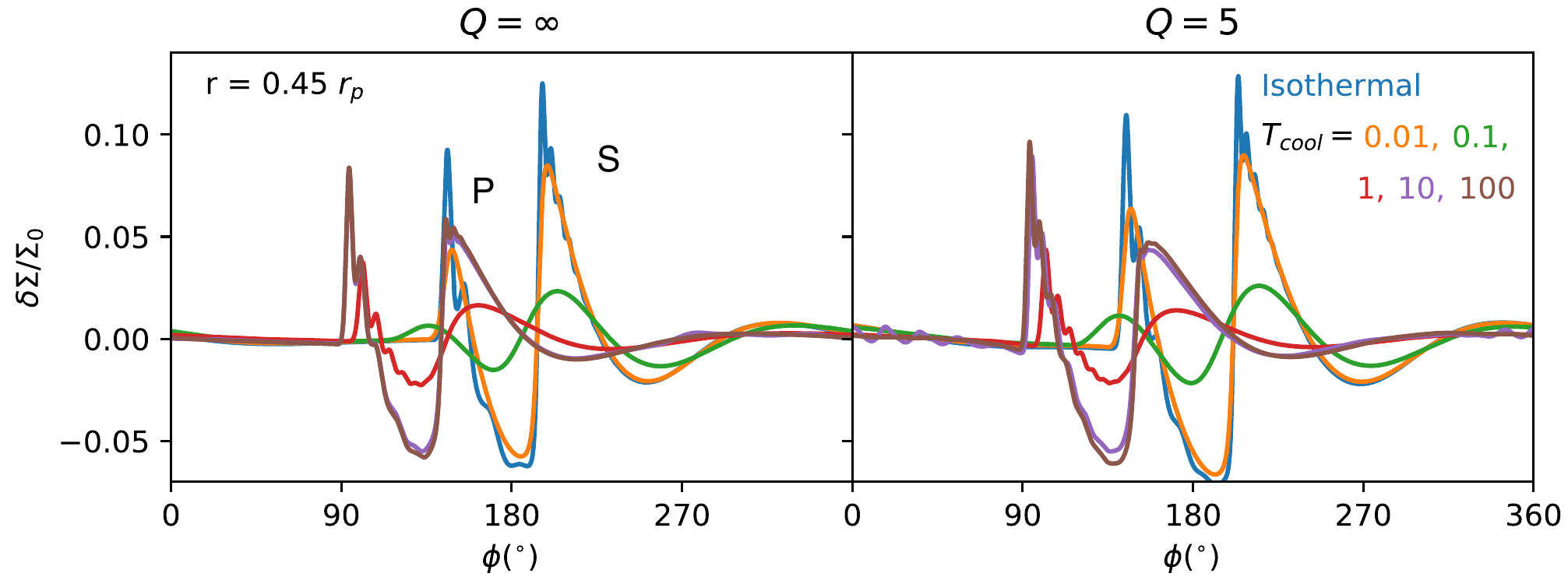}  
\includegraphics[width=\linewidth]{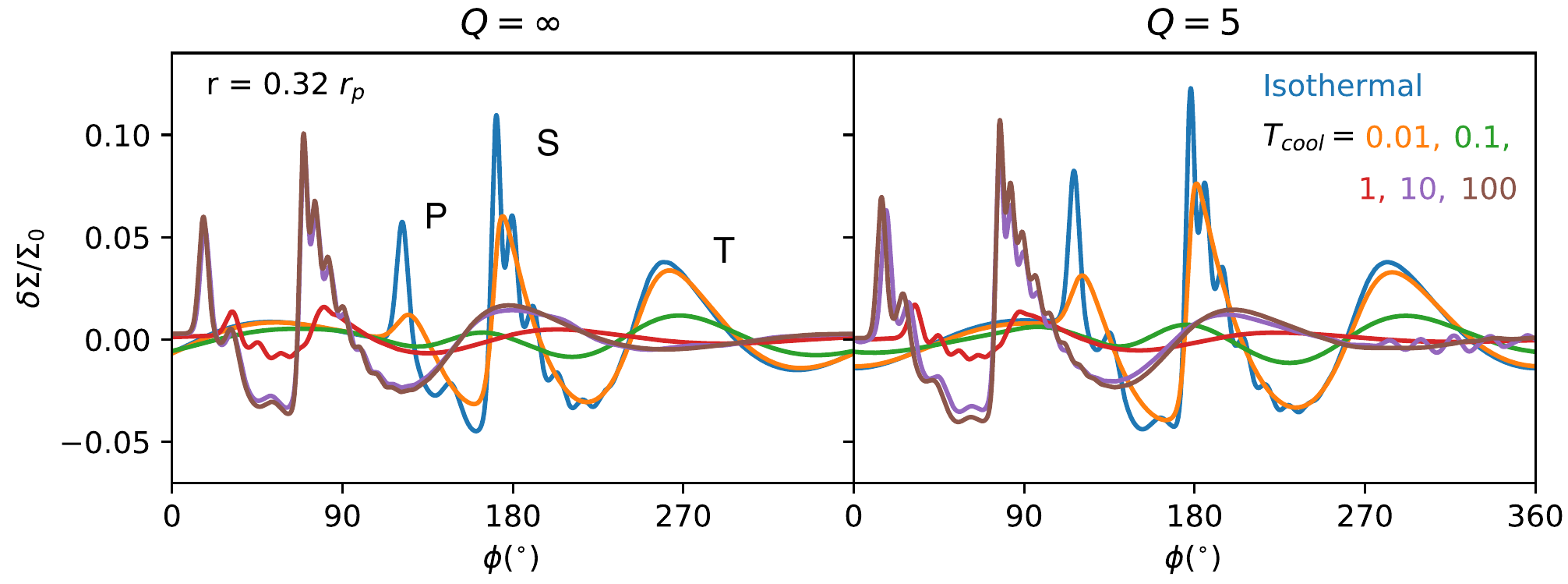} 
\caption{The density perturbations along $r= 0.45\ r_p$ (upper panels) and $r = 0.32\ r_p$ (lower panels) for \texttt{B17AD} (left panels) and \texttt{B17ADSG} (right panels) disks at t = $40\ t_p$. The positions of the primary, secondary and tertiary spirals for small $T_{cool}$ are marked as ``P'', ``S'' and ``T'' on the left panels.
\label{fig:ADSGradialcut}}
\end{figure*}

The change of the amplitude and position of the spiral arms with respect to the cooling time can be seen more clearly in 1D plots.  Figure \ref{fig:ADSGradialcut} (left panels) shows the density perturbations vs. azimuthal angle ($\phi$) cut at $r = 0.45\ r_p$ and $r = 0.32\ r_p$. The density perturbations due to the primary spiral and the secondary spiral are marked as ``P'' and ``S'' on the figure. The one due to the tertiary spiral is also marked as ``T'' in the $r = 0.32\ r_p$ plot. The positions of the primary, secondary, and tertiary spirals in $T_{cool}$ = 0.01 disks are similar to those in the isothermal disk. At $r = 0.45\ r_p$, the amplitude of the spiral in the $T_{cool}$ = 0.01 disk is slightly lower than that in the isothermal disk.

There is a significant change of the spiral's amplitude when $T_{cool}$ increases from $\lesssim$ 0.01 to 0.1. The spiral's amplitude becomes much smaller at $T_{cool}=0.1$ with a slight shift of the primary spiral position. The tertiary spiral almost disappears at $r=0.32\ r_p$, but we can still see its presence with a amplitude of $\sim 1\%$. But the biggest change of the spiral's position occurs when $T_{cool}$ increases from 0.1 to 1. At $r = 0.45\ r_p$, the primary spiral shifts $\sim45^o$ between $T_{cool}=0.1$ to $T_{cool}=1$ cases, although the amplitude does not change much. The tertiary spiral at $r=0.32\ r_p$ is now the hump from $160^{\circ}$ to $250^{\circ}$ with a amplitude of $\sim 0.5\%$.   At $T_{cool} = 10$, the amplitude of the spiral becomes much higher. The positions are very similar to those in $T_{cool} = 1$. The tertiary spiral also becomes stronger. But the spirals are still weaker than those in the isothermal disk.  From $T_{cool}= 10$ to $T_{cool} = 100$ disks, the perturbations seem to converge to the limit of an adiabatic disk. Overall, from isothermal disks ($T_{cool} \lesssim 0.01$) to adiabatic disks ($T_{cool} \gtrsim 10$) the azimuthal angles of the spirals shift to smaller values by $\sim 50 ^{\circ}$ at $r = 0.45\ r_p$. This is consistent with the finding in 2D plots that the spirals become less tightly wound as $T_{cool}$ increases \footnote{Aside from the interesting behavior at $T_{cool}\sim$ 1, we want to
note that at two extremes, the spiral at the adiabatic limit is still weaker than that at the locally isothermal limit as expected. This is related to the increase of the effective sound speed with a longer cooling time. When the sound speed increases, the thermal mass increases and ratio between the planet mass and the thermal mass decreases so that the planet is less capable of exciting spiral waves.}.

\begin{figure} 
\includegraphics[width=\linewidth]{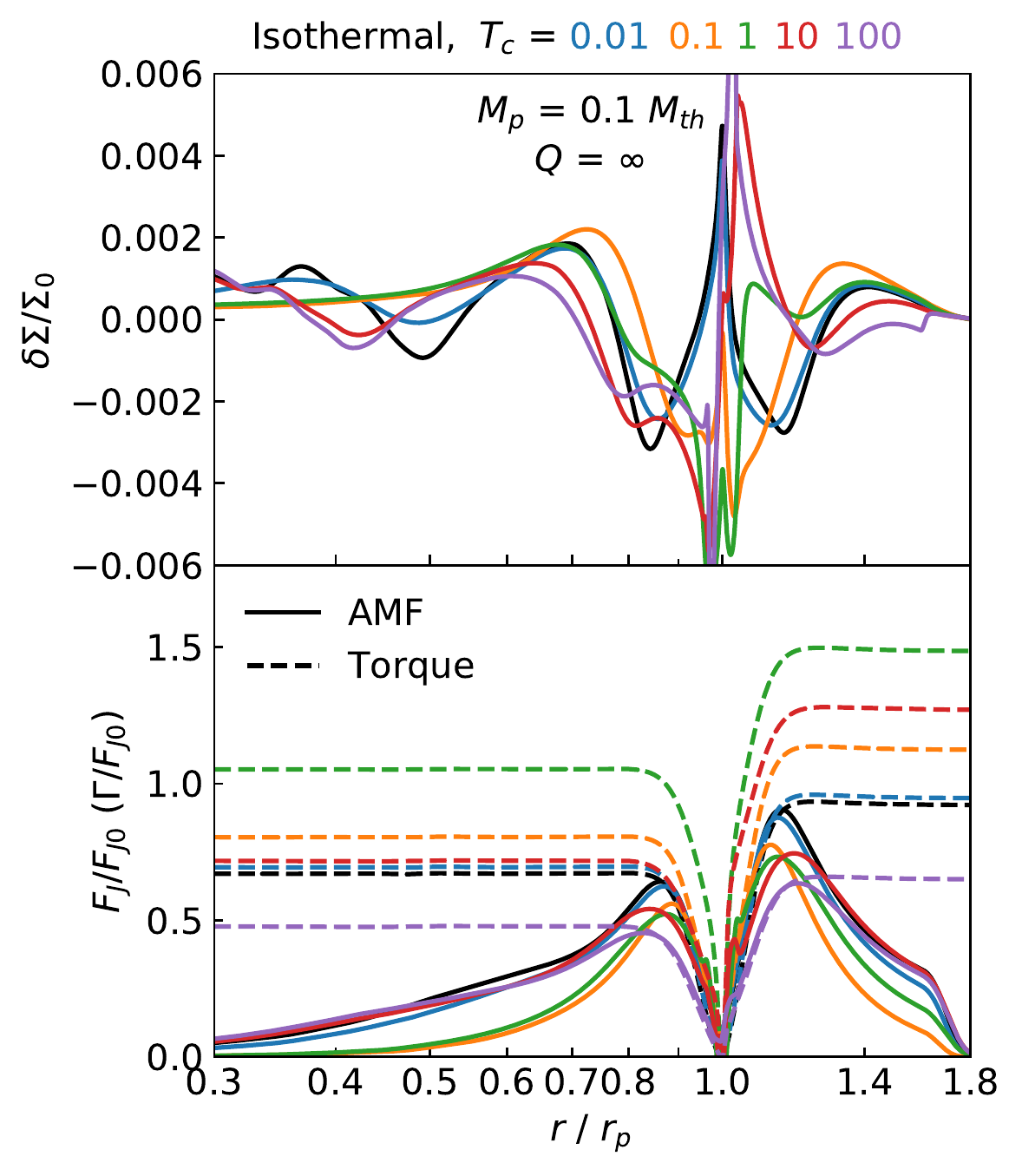}  
\caption{Top Panel: The azimuthally averaged density perturbations in Figure \ref{fig:adiabatic2Dgasdens}, upper panels.   Bottom Panel: The normalized AMF (solid curves) and torque (dashed curves) for different cooling times, $T_{cool}$. The black curves represent isothermal disk, whereas blue, orange, green, red and purple curves represent adiabatic disks with $T_{cool}$ = 0.01, 0.1, 1, 10 and 100, respectively. The changes of the AMF distribution should be responsible for the change of the density perturbations.
\label{fig:adiabaticdensAMF40orbits}}
\end{figure}

Since spirals are directly related to gap opening, we want to study how the properties of the induced gaps are related to  radiative cooling.
The top panel of Figure \ref{fig:adiabaticdensAMF40orbits} shows the azimuathlly averaged density perturbations of disks with different cooling times. $T_{cool}$ = 0.01 disk has a similar profile to the isothermal disk, but with smaller amplitudes. The secondary gap at $T_{cool}$ = 0.1 and 1 disappears (the disappearance of the gap in gas does not necessarily lead to that in dust; see Section \ref{sec:AS209}). It reappears at $T_{cool}$ = 10, but the position changes from 0.5 $r_p$ to 0.4 $r_p$ and the amplitude becomes smaller. Since spirals carve gaps in disks, the inward shift of the secondary gap position is due to the more open secondary spiral for higher $T_{cool}$. On the other hand, the non-monotonically transition---the trend of high, low and high amplitudes of the secondary gap depth from the fast cooling to the slow cooling cases--- is due to weaker spirals at intermediate $T_{cool}$. 
Overall, for the intermediate massive planet ($M_p \sim 0.1\ M_{th}$), disks with $T_{cool} \lesssim 0.01$ can be treated as isothermal disks and disks with $T_{cool} \gtrsim 10$ can be seen as purely adiabatic disks, whereas the disks undergo a sharp transition between these limits, characterized by low amplitude spirals and shallow secondary gaps. Note that this $T_{cool} \lesssim  0.01$ criterion is more stringent with a low mass planet in the linear regime. As shown in Figure \ref{fig:BZ18_1dDensAMF}, only when $T_{cool} = 10^{-4}$ the AMF is close to the isothermal case with $M_p = 0.01\ M_{th}$. Both the temperature gradient (T $\propto r^{-0.5}$) and the linear damping contribute to the AMF profile, but the linear damping is the dominant effect here.

Since such conditions with intermediate cooling times are very likely to occur in realistic protoplanetary disks (see Section \ref{sec:theory}),  we are interested in whether these much shallower secondary gaps can still trap enough dust particles to be observable. Examining the previous models in  \citet{zhang18}, we find the current set of simulations has very similar setup to the $h/r = 0.07$, $\alpha = 10^{-4}$, $M_p/M_* = 11 M_\oplus/M_\odot$ model in \citet{zhang18} (the upper-left plots in 2D figures therein), despite the $\alpha$ is lower here. However, the secondary gap cannot be found in any dust configurations in that model, even though those simulations assume the locally isothermal EoS. Thus, we conclude that the models being analyzed in this subsection with such low planet mass is not suitable for testing the observability of the secondary gap. To that end, we run simulations with higher planet mass, 1 $M_{th}$, close to the setup of the AS 209 simulation, $\alpha$ varying model \citep{zhang18}. We direct the discussion to Section \ref{sec:AS209}.

We also plot the AMF (in solid curves) and the integrated torque (in dashed curves) of these disks in the lower panel of Figure \ref{fig:adiabaticdensAMF40orbits}. The torque is integrated from the planet's position towards either side of the disk. The AMF decreases monotonously close to the planet ($|r-r_p| \lesssim 0.1\ r_p$) due to the torque cutoff. The AMF far away from the planet (e.g., at r=0.3 $r_p$ and 1.5 $r_p$) decreases to the lowest values at $T_{cool} =$ 0.1 and 1 and increases back to the isothermal values at $T_{cool} =$ 10 and 100. On the other hand, from the fast cooling (and isothermal) to the slow cooling cases, the torque first increases and reaches the maximum at $T_{cool}$ = 1 and then decreases and reaches the minimum at $T_{cool}$ = 100. The AMF and torque at several $h$ away from the planet  (e.g., at $r$ = 0.85 $r_p$ or 1.2 $r_p$) have similar values in both the very small (e.g., $T_{cool}=0.01$) and very large (e.g., $T_{cool}=100$) $T_{cool}$ cases. However, the torque there is much higher than the AMF at $T_{cool}$= 0.1, 1 and 10. This indicates that, for intermediate cooling cases, the waves fail to launch or are damped right after they are launched, indicating radiative cooling is important for these cases. Note that the normalization (characteristic value, $F_{J0}$) should have varied with different cooling times, since $F_{J0} \propto h_p^{-3}$. In adiabatic EoS, the sound speed, $c_s = \big(\gamma\frac{dp}{d\rho}\big)^{1/2}$ and the scale height $h \propto c_s \propto \gamma^{1/2}$. Thus, $F_J \propto \gamma^{-3/2}$. The effective normalization is 0.6 times the $F_{J0}$ in Equation \ref{eq:FJ0} at the slow cooling limit when $\gamma = 1.4$. This value should be in between 1 and 0.6 in transition from small to large $T_{cool}$. Nevertheless, we just use $F_{J0}$ as Equation \ref{eq:FJ0} and compare the absolute values of the AMF.
We will have more detailed discussions in Section \ref{sec:AMFvs}.

\subsection{Disks with both self-gravity and radiative cooling}
We also explore a situation that both self-gravity and radiative cooling are included, even though we do not seek to carry out a parameter space study. The parameters we choose are $Q=5$, \texttt{B17ADSG} disks with various $T_{cool}$. The density perturbations at two radii are shown in Figure \ref{fig:ADSGradialcut}'s right panels. The profiles of the spirals are very similar to the non-selfgravitating ones on the left panels. All the trends found in Section \ref{sec:resultsAD} still apply. At the same cooling time, the perturbation amplitudes become stronger and the spirals become tighter if $Q$ becomes smaller, which are consistent with the results in Section \ref{sec:resultSG}. This means that when both of these two physical processes (disk self-gravity and radiative cooling) are operating, they play similar roles as those that they play individually, at least for low mass planets. From the magnitudes of the changes due to these two processes, we can conclude that when the disk is not too massive ($Q \gtrsim$ 5), the effects of the self-gravity on the disks are less important than those of the cooling.

\section{Discussion}\label{sec:discussion}

\subsection{AMF vs. $Q$ and \texorpdfstring{$T_{cool}$}{}} \label{sec:AMFvs}

\begin{figure*} 
\includegraphics[width=\linewidth]{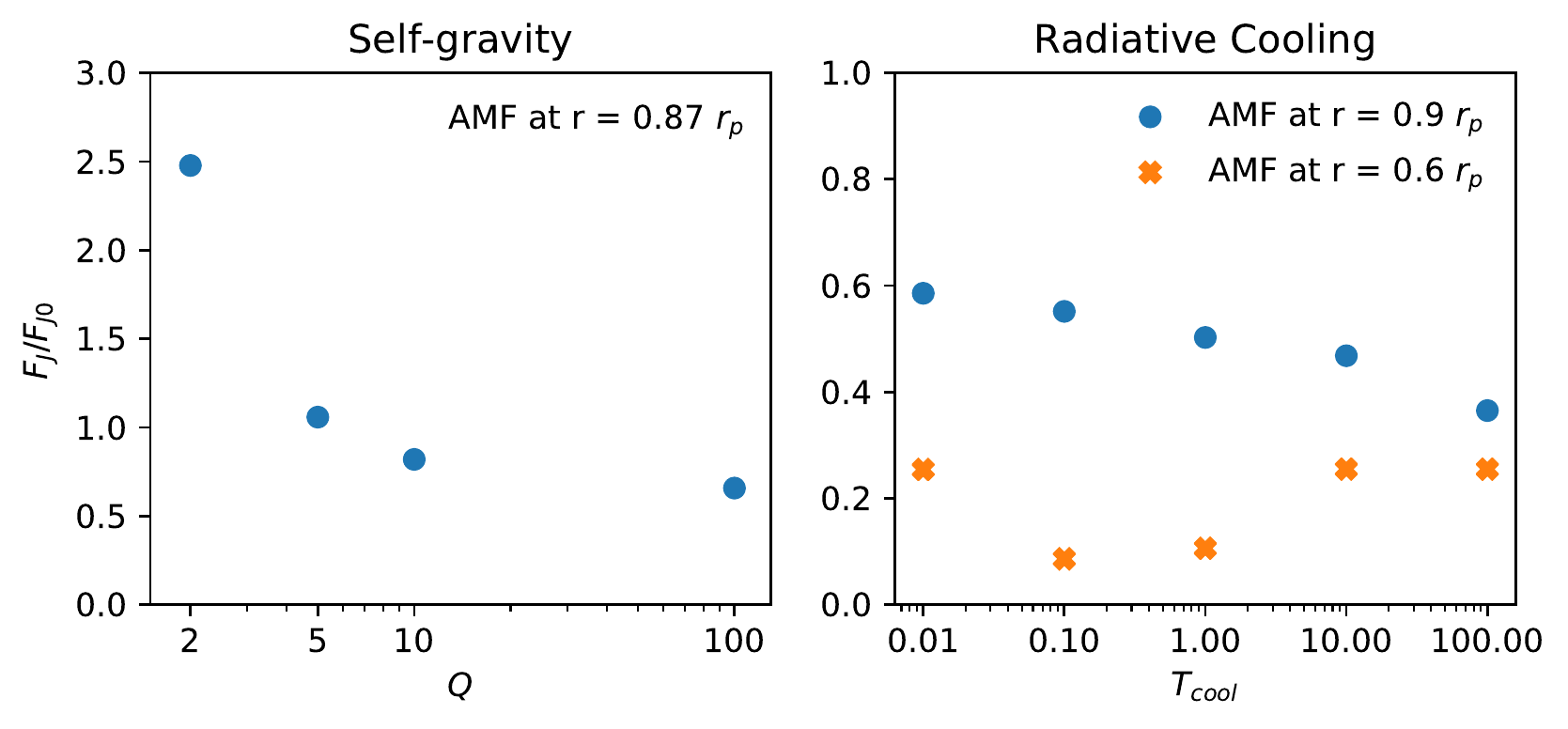} 
\caption{AMF vs. $Q$ (left) and AMF vs. $T_{cool}$ (right). The AMF is measured at $r= 0.87\ r_p$ for the \texttt{SG} runs and at $r= 0.6\ r_p$ (marked as ``X''s) and $0.9\ r_p$ (marked as dots) for the \texttt{AD} runs.
\label{fig:FJvsQandTcool}}
\end{figure*}

To quantify the relationships between AMF vs. $Q$ and AMF vs. $T_{cool}$, we measure the AMF at given positions in Figure \ref{fig:BZHAMF50orbits} and Figure \ref{fig:adiabaticdensAMF40orbits} for \texttt{B17SG} and \texttt{B17AD} disks with $M_p$ = 0.1 $M_{th}$. We plot the AMF as a function of $Q$ (the left panel) and $T_{cool}$ (the right panel)  in Figure \ref{fig:FJvsQandTcool}. The AMF is measured at 0.87 $r_p$ for \texttt{B17SG} simulations, where it reaches the maximum at the inner disk. The AMF increases as the disk becomes more massive (with a smaller $Q$). This increase becomes quite fast when $Q$ is approaching 2. The advective AMF in the $Q=2$ simulation is almost 5 times the AMF in the $Q=100$ simulation, which is roughly consistent with the analytical expectation (Equation \ref{eq:fQGT80}). 

As for the radiative cooling, we measure the AMF at 0.9 $r_p$ and 0.6 $r_p$. At the position very close to the planet (e.g., 0.9 $r_p$), the AMF decreases as the cooling time increases. As mentioned in Section \ref{sec:resultsAD}, this is most likely due to the transition of the effective $\gamma$ (that enters the sound speed) from 1 to 1.4. The ratio between the AMF at $T_{cool} = 100$ and that at $T_{cool} = 0.01$ is $\sim$ 0.6, which is very close to the expected change due to our definition of $F_{J0}$. However, if the AMF is measured farther away from the planet (e.g., 0.6 $r_p$), it first decreases and then increases as $T_{cool}$ increases, reaching the minima at 0.1 and 1 $T_{cool}$.  The decrease of the AMF with the distance away the planet is due to the wave damping as the wave propagates. When the wave propagates to $0.6\ r_p$, the damping is the strongest at $T_{cool} \sim 0.1-1$, implying that radiative cooling plays an important role on the wave damping/dissipation when $T_{cool}\lesssim 1$. We are expecting a stronger wave damping with a smaller $T_{cool}$. However, when $T_{cool}$ becomes very small and approaches the isothermal limit, the temperature of the wave becomes the background disk temperature, and shock damping in this case behaves similar to the shock damping in the adiabatic limit \citep{goodman2001, dong2011}. 
At the position that the secondary spiral should form, the wave becomes already too weak to perturb the gas for the $T_{cool}=0.1-1$ cases. Thus, the secondary gap can even disappear at these intermediate cooling times.  

\subsection{Implications for observations} \label{sec:implications}
To quantify the properties of the primary gap, we measure the gap width and depth in the gas density perturbation maps for \texttt{B17SG} and \texttt{B17AD}, $M_p$ = 0.1 $M_{th}$, using the definitions in  \citet{zhang18}. Different from  \citet{zhang18}, we only measure the gaps at $\sim 50$ orbits and from the gas density perturbation profiles, because we do not run the simulations for very long time and the planet mass is still too small to open significant gaps in gas and dust. However, these measurements suffice to demonstrate the trend. 

\begin{figure} 
\includegraphics[width=\linewidth]{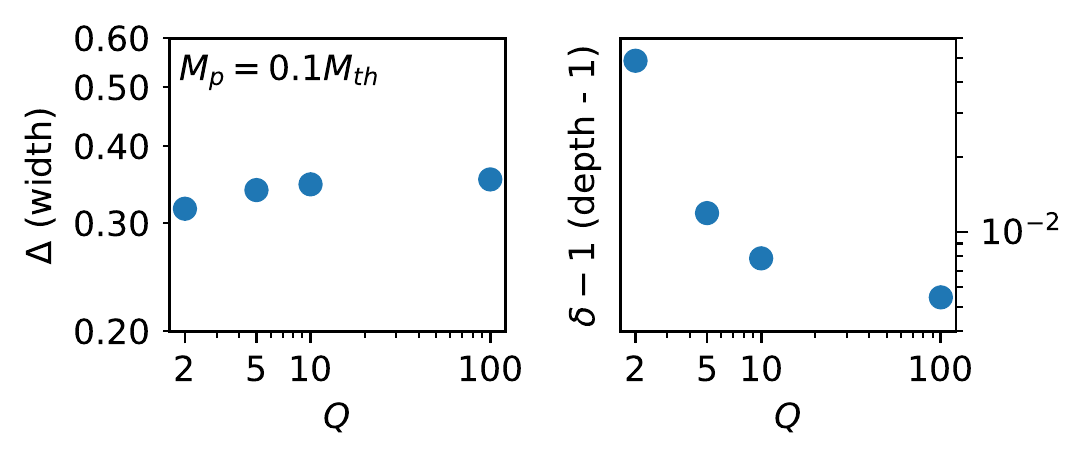} 
\includegraphics[width=\linewidth]{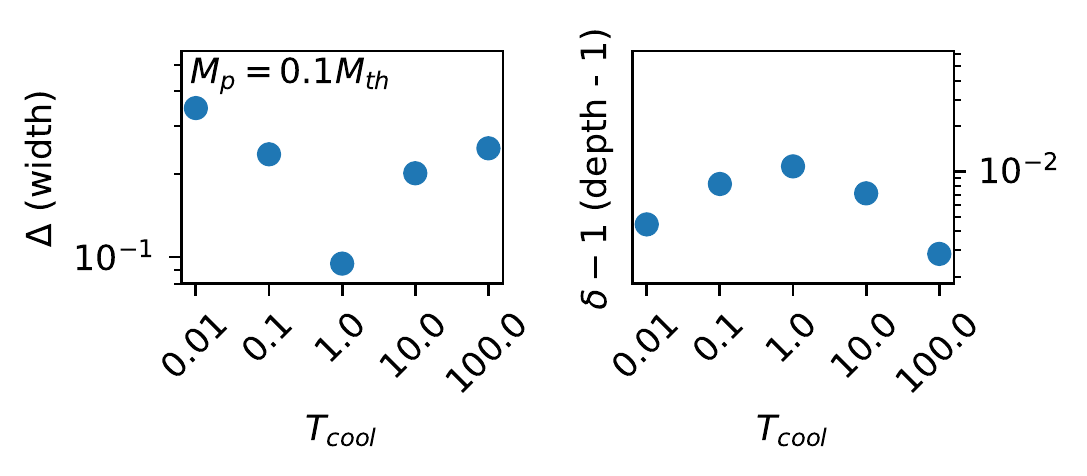} 
\caption{Gap width/depth vs. $Q$ (top) and Gap width/depth vs. $T_{cool}$ (bottom). $\Delta$ and $\delta$ - 1 are proportional to $M_p$. The range for the gap (depth - 1) is 5 times than that in the width. Under this factor, the y-axes on the left and right panels have approximately the same scaling to $M_p$. Thus, the y-axes of the left and right panels can be directly taken as the planet mass in an arbitrary unit.
\label{fig:WidthdepthvsQandTcool}}
\end{figure}

In the upper panels of Figure \ref{fig:WidthdepthvsQandTcool}, the gap width - $Q$ relation is shown on the left panel, whereas the depth - $Q$ relation is shown on the right panel. Both the x and y axes are in logarithmic scales. Since the exponents in the gap depth relations are $\sim 5$ times larger than that of gap width relations (see Table 1 and 2 in  \citealt{zhang18} for details), the plotted y-axis range for the depth ($\delta -1$) is 5 times of the range for the width ($\Delta$). In this way the relative magnitude read on both y-axes is roughly proportional to the planet mass ratio.  As $Q$ changes to lower values, the gap depth has more significant changes than the width. The planet mass inferred from the gap depth without considering disk self-gravity will be overestimated. Adopting common exponents 0.25 and 1.25 for the width-$M_p$ and depth-$M_p$ relations (more precisely, width-$K'$ and depth-$K$ in \citealt{zhang18}), the ratio of the inferred masses between $Q=2$ and $Q=100$ disks is 1.5 using the width-$M_p$ relation, and 6 using the depth-$M_p$ relation. Thus, if one wants to infer the planet mass, the gap width is a better observable than the gap depth since it is less sensitive to the disk mass. 

The lower panels of Figure \ref{fig:WidthdepthvsQandTcool} show the width and depth as functions of the $T_{cool}$. The gap width and depth are not very well defined due to the irregular gap shape. We try to neglect the horseshoe region around $r_p$ when we measure the gap width. It is better to use the gap depth as a proxy for the planet mass. If one uses the gap width to measure the planet mass in locally isothermal simulations, the planet mass will be underestimated if $T_{cool} \gtrsim 0.01$.

Note that these trends found in the low planet mass regime do not necessarily apply to the high planet mass regime, as we will discuss in the next subsection, where we find much less change in the gap shape for various cooling times.

\subsection{AS 209: cooling in the high planet mass regime} \label{sec:AS209}

\begin{figure} 
\includegraphics[width=\linewidth]{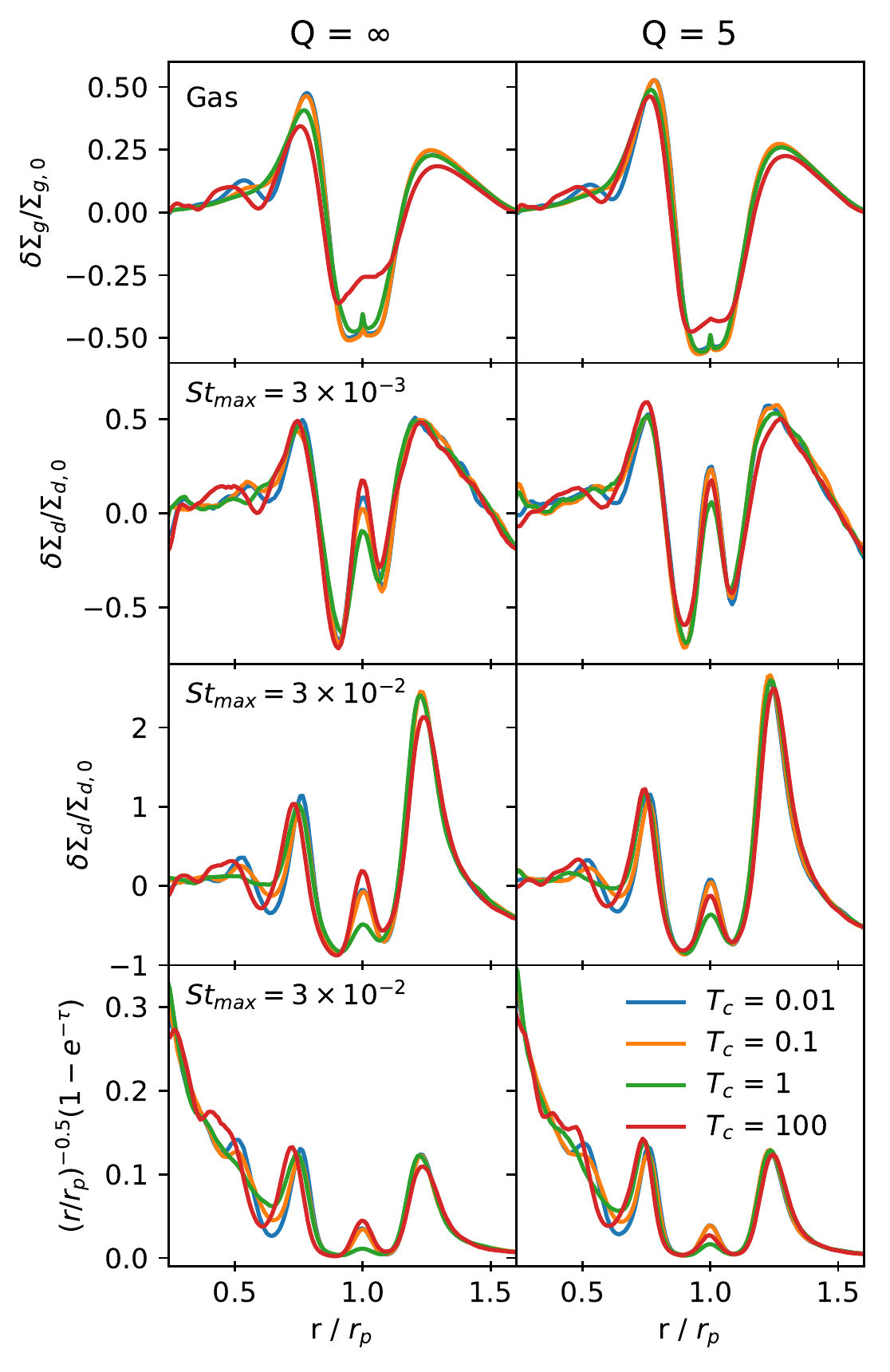} 
\caption{Azimuthally averaged density perturbations and brightness temperature of \texttt{AS209AD} and \texttt{AS209ADSG} at 400 $t_p$. From top to bottom, they are the density perturbations of the gas, dusts with $St_{max} = 3\times 10^{-3}$ and $St_{max} = 3\times 10^{-2}$. The bottom panels show the brightness temperature of the $St_{max} = 3\times 10^{-2}$ ($a_{max} = 0.5\ \mathrm{mm}$) dust models. The left panels show the cases without self-gravity, whereas the right panels show the cases with $Q=5$. Blue, orange, green and red curves represent $T_{cool}$ = 0.01, 0.1, 1 and 100, respectively. The $T_{cool} = 0.01$ curves on the left panels are almost identical to the locally isothermal cases in \citet{zhang18} (the slight difference is mainly due to the different setups in resolution, temperature profile, indirect term and smoothing length).}
\label{fig:AS209}
\end{figure}

We have two goals in this subsection. One is to explore the effects of radiative cooling on the gap properties when the planet mass is relatively large reaching the thermal mass. Meanwhile, we decide to choose a model that resembles a realistic disk so that we are able to constrain its surface density via estimating its $T_{cool}$. Of all the DSHARP disks, AS 209 features many intricate gaps and rings \citep{guzman18}. Using simple (without radiative cooling and self-gravity) hydrodynamical simulations, \citet{zhang18} find an excellent match to the AS 209 disk with a model whose $(h/r)_p = 0.05$, $\alpha=3\times 10^{-4} (r/r_p)^2$ and $M_p/M_* = 10^{-4}$, or 1 $M_{th}$\footnote{We also run simulations with constant $\alpha$ as the model (b) in \citet{zhang18}. Same as previous models, they reproduce AS 209's 1D profile, but are non-axisymmetric in 2D. We do not see more differences due to the viscosity profile.}. The dust size distribution is a power law $n(a) \propto a^{-3.5}$, $a_{max}$ = 0.68 mm, and the gas surface density $\Sigma_{g,0} = 6.4\ \mathrm{g\ cm^{-1}}$. This model can explain up to five gaps in the disk, not only their locations, but also the intensity. Since we have found that self-gravity and radiative cooling have the potential to change the gap shape and the secondary gap position in the low planet mass regime, we would like to explore how the planet mass for AS 209  will change considering these two physical processes. 

Figure \ref{fig:AS209} shows the azimuthally averaged density perturbation and  brightness temperature of the \texttt{AS209} models with different cooling times ($T_{cool}$ = 0.01, 0.1, 1 and 100) and  different disk masses ($Q = \infty$ and 5) at 400 $t_p$. We choose not to show $T_{cool}$ = 10 case to avoid crowdedness of the figure, but its feature is similar to the $T_{cool}$ = 100 case with shallower gaps in dust. The left panels show the $Q = \infty$ cases, whereas the right panels show the $Q = $ 5 cases. Blue, orange, green and red curves represent $T_{cool}$ = 0.01, 1, 10, and 100 cases, respectively. From the top to bottom panels, it shows the density perturbations of the gas,  the dust with $St_{max} = 3 \times 10^{-3}$, $St_{max} = 3 \times 10^{-2}$, and the brightness temperature of the $St_{max} = 3 \times 10^{-2}$ models. We assume that the temperature in the disk is proportional to $r^{-0.5}$ in the last row and the procedure to obtain the optical depth is detailed in \citet{zhang18}. As the $T_{cool}$ becomes larger, the perturbation at the major gap (where the planet locates) becomes shallower in gas. However, this is not the case in dust. The gap width and depth are almost the same across different cooling times. This result is different from that in the low planet mass regime, where the gap shapes are quite different across different cooling times. For dust profiles, the horseshoe region has the smallest density for $T_{cool}$ = 1. The secondary gap disappears for $T_{cool} = 1$ in gas, but it is still noticeable in dust for $St_{max}=3\times 10^{-2}$, with a lower amplitude than $T_{cool}$ = 0.01, 0.1, and 100 cases. 
This indicates that, due to the gaseous bump at the inner edge of the primary gap, dust particles can still be trapped there, forming the secondary gap in dust even if the secondary gap disappears in the $T_{cool}=1$ case. The location of the secondary gap shifts from $\sim 0.65\ r_p$ to $\sim 0.6\ r_p$ when $T_{cool}$ increases. The tertiary gap is also present in $T_{cool} = 0.01$ and 100 cases. It moves inwards from $\sim 0.43\ r_p$ to $\sim 0.37\ r_p$. Although the tertiary gap is not present in the $T_{cool} = 1$ case, a change of slope can been seen at $\sim 0.4\ r_p$ (this is more obvious in $Q=5$ case). The $Q = 5$ disks are similar to the $Q=\infty$ disks but with slightly stronger perturbations. The differences between different $T_{cool}$ curves are slightly smaller in $Q=5$ disks than those in $Q=\infty$ disks.

Note that aside from the EoS, the simulation setups in this subsection are different from those in \citet{zhang18} in several aspects (e.g., the resolution, domain size, temperature profile, indrect term, smoothing length, and planet growing time). Thus, the gap shape and position are expected to be slightly different from the simulations in \citet{zhang18}. Also, as time evolves, the position of the secondary and tertiary gaps will move slightly inwards. Thus, we do not try to compare these profiles with  AS 209 observations. Instead, we use this example to demonstrate the observable differences due to different cooling times.

We also want to emphasize that without the self-gravity, the location of the planet and the gaseous surface density from the simulation can be scaled to any value. However, when the disk self-gravity is included, the gas density and the location of the planet are related. The relation is $\Sigma_{g,0} = (M_*/r_p^2)(h/r)_p/(\pi Q_p)$. Taking $M_*$ = 0.83 $M_\odot$, $r_p = 99\ \mathrm{au}$, $Q_p = 5$ and $(h/r)_p = 0.05$, we get $\Sigma_{g,0}$ = 2.4 $\mathrm{g\ cm^{-2}}$. 
In \citet{zhang18},
the best fit model has $Q \approx 2$ at the planet's position in the initial condition.

Assuming the gaps at 24 au, 35 au, and 61 au in AS 209 are related to the planet at 99 au, we can use these gaps to constrain the AS 209 disk mass. Using Equation \ref{eq:Tcool}, and adopting the stellar luminosity as 1.4 $L_\odot$ and the mass as 0.87 $M_\odot$ from \citet{andrews18b}, $\phi = 0.02$, $T_f = 20$ K, and $\kappa_R = 0.21\ \mathrm{cm^2 g^{-1}}$, we have
\begin{equation}
    \begin{split}
     T_{cool} = 0.02 \Bigg[1+ 0.01\Big(\frac{\Sigma}{\mathrm{1\ g\ cm^{-2}}} \Big)^2 \Bigg]
     \begin{cases} 
        2.5 & r \leq 46\ \mathrm{au} \\
        \Big(\frac{r}{99\ \mathrm{au}}\Big)^{-3/2} & r > 46\ \mathrm{au} \\
      \end{cases}\,.
   \end{split}
\end{equation}
To produce an observable tertiary gap, we need low or high cooling times ($T_{cool} \lesssim 0.1$ or $T_{cool} \gtrsim 10$). However, $T_{cool} \gtrsim 10$ requires $\Sigma \gtrsim 200\ \mathrm{g\ cm^{-2}}$, which is impossible since $Q$ would be $\ll 1$.  If $T_{cool}\lesssim 0.1$, we can constrain $\Sigma \lesssim 20 \mathrm{\ g\ cm^{-2}}$. While it is an independent constraint, this does not give tighter constraint on the density for this disk, since from gravitational instability criterion (Equation \ref{eq:toomreQ}), the disk will become unstable if $\Sigma_{g,0} > 12\ \mathrm{g\ cm^{-2}}$ at 99 au. Nevertheless, this constraint from the disk cooling is consistent with that from the disk instability. 

The $Q$ = 5 models with low $T_{cool}$ might be plausible models for the AS 209 disk, since their density at 99 au is allowed given the constraints derived above. However, with the  density $\Sigma_{g,0}$ = 2.4 $\mathrm{g\ cm^{-2}}$ from the $Q=5$ disk, we cannot restore the intensity measured from the DSHARP observation. It is much lower than that from the observation. Only $\Sigma_{g,0} \gtrsim 6\ \mathrm{g\ cm^{-2}}$ can explain the intensity. This means that either the actual opacity is higher than what we adopt (e.g., the dust-to-gas mass ratio is larger) or the disk has even lower $Q$. If the latter is true, the $T_{cool}$ can be larger (but still $< 1$), since the low $Q$ that is close to unity can produce much stronger perturbations, compensating for the weaker perturbations due to the larger value of the $T_{cool}$. Interestingly, \citet{powell19} estimate the gas surface density $\Sigma \approx 10\ \mathrm{g\ cm^{-2}}$ at 100 au, using the ``dust line'' method. This high surface density is actually compatible and consistent with the constraints from disk self-gravity and radiative cooling for the AS 209 disk. If this is the case, it indicates that the outer disk of the AS 209 disk is marginally gravitational stable.

\subsection{Long time evolution for self-gravitating disks with planets} \label{sec:unstable}
\begin{figure*} 
\includegraphics[width=\linewidth]{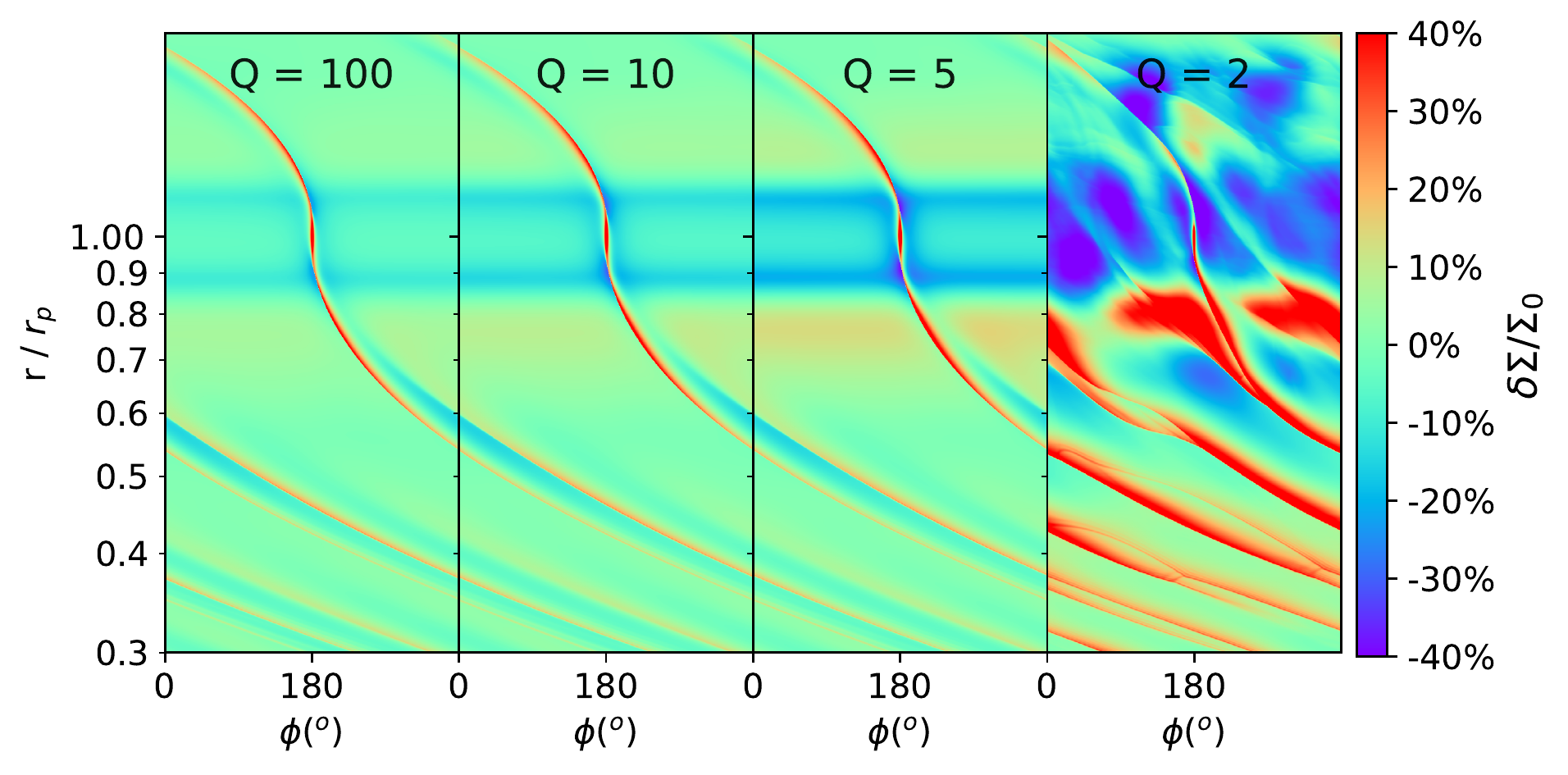} 
\includegraphics[width=\linewidth]{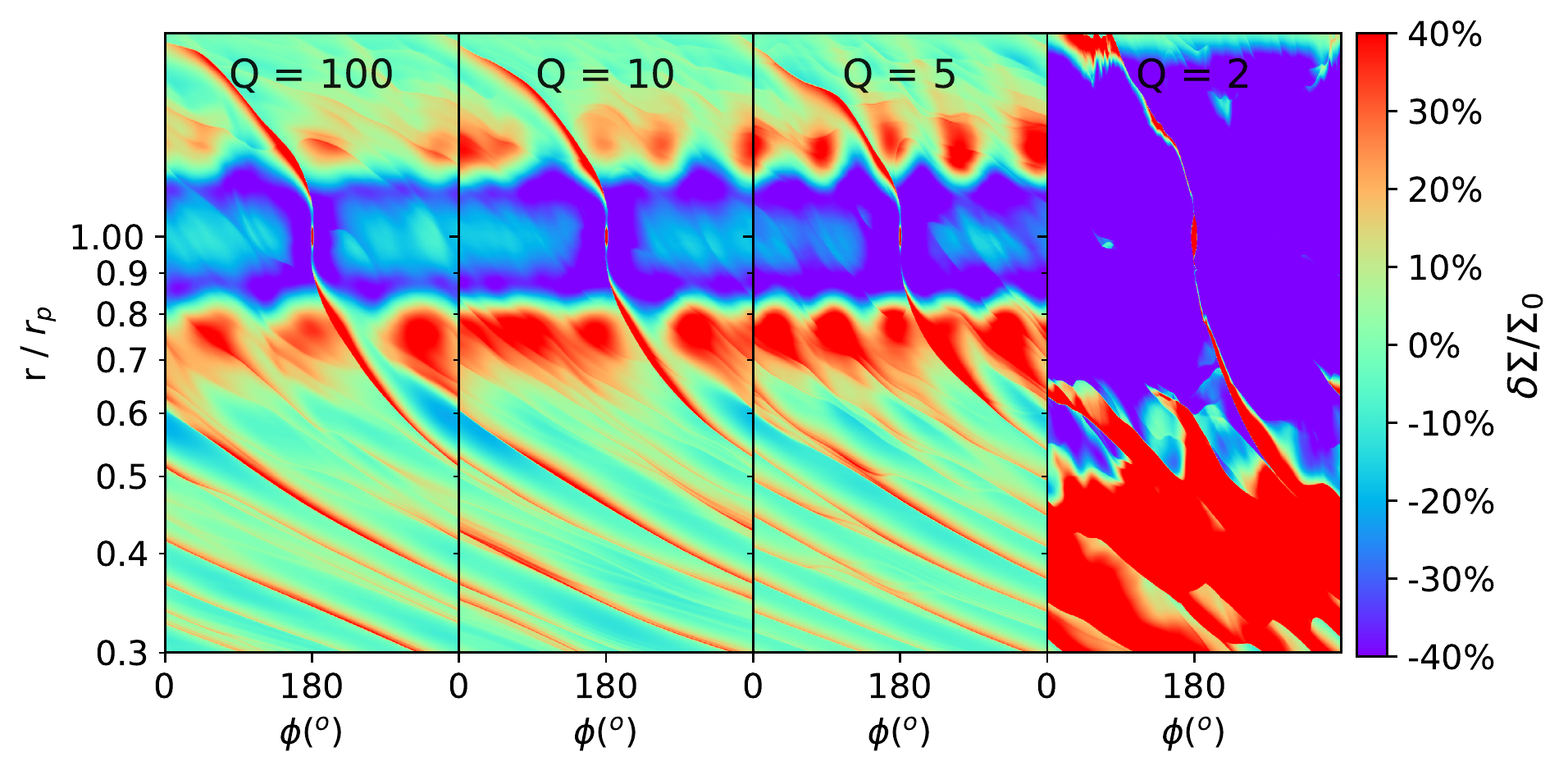} 
\caption{The density perturbations of \texttt{B17SG} disks with $M_p$ = 0.3 and 1.0 $M_{th}$ (top and bottom) at $t = 100\ t_p$. The disk becomes unstable at $Q = 2$. The lower panel shows that self-gravity can suppress small $m$ vortices, so that the number of the vortices increases as $Q$ becomes smaller. 
\label{fig:BZH17SG2dGas_100orbits}}
\end{figure*}

On the other hand, in our long timescale simulations, we find that the self-gravitating disks with planets tend to become unstable at later times. This happens earlier if the $Q$ is small or if the planet mass is high. Figure \ref{fig:BZH17SG2dGas_100orbits} shows the density perturbations of \texttt{B17SG} disks when $M_p$ = 0.3 and 1.0 $M_{th}$ (top and bottom) at $t = 100\ t_p$. $Q=2$ disks have become unstable at this point. This is why low $Q$ and high $M_p$ curves are missing in Figure \ref{fig:BZHgasdustdens1d}. The disk would become more stable if the radiative cooling is included. For instance, the \texttt{AS209ADSG} simulations discussed in Section \ref{sec:AS209} can evolve for very long time without any sign of instability even though their $Q=5$ and $M_p = 1\ M_{th}$. Another interesting behavior in the lower panels is that as the disk becomes more massive, the number of vortices at the gap edge becomes larger. This is consistent with the simulation results in \citet{lin12, zhu16}, as self-gravity can suppress large-scale (small $m$) vortices produced by Rossby wave instability \citep{lovelace99}.

\section{Conclusions}\label{sec:conclusion}
We run two dimensional hydrodynamical simulations to explore the effects of self-gravity and radiative cooling on the gap and spiral formation in protoplanetary disks. We explore these effects with different Toomre $Q$ parameters and the dimensionless cooling times $T_{cool}$ for low mass planets. As the disk becomes more massive (smaller $Q$), we find:
\begin{itemize}
    \item The spirals become slightly more tightly-wound, especially when $Q$ $\sim$ 2, which is consistent with the linear theory.
    \item The spirals become stronger and the primary and secondary gaps become deeper. This is due to the higher AMF in more massive disks, which is consistent with the linear theory. The advective AMF in the $Q=2$ disk is almost 5 times the AMF in the $Q=100$ disk.
    \item If $Q \gtrsim 2$, the secondary gap's position does not change, despite gaps being deeper.
 \end{itemize}
 
In disks with radiative cooling, we find:
\begin{itemize}    
    \item Even with a relatively short $T_{cool}$ = 0.01, the spiral waves cannot pick up AMF from the background disk when they are propagating to hotter regions, different from locally isothermal disks.
    \item The spirals become less tightly-wound (more open) as the cooling time increases, due to the increase of the disk's effective sound speed. The spiral's openness changes dramatically from disks with $T_{cool}=$ 0.1 to $T_{cool}=$ 1.   
    \item The spirals become weaker as the cooling time increases from a small value to $1/\Omega$, and then start to become stronger as the cooling time increases. There is a significant change of the spiral's amplitude when $T_{cool}$ increases from $\lesssim$ 0.01 to 0.1.
    \item The AMF of the wave dissipates the fastest during its propagation when $T_{cool}\sim 0.1$ to 1, implying that radiative damping may be important.
    \item  With weaker spirals, the secondary gaps due to low mass planets disappear when $T_{cool}=0.1$ to 1.
    \item If the secondary gap is present, its position moves from $\sim$ 0.5 $r_p$ to $\sim$ 0.4 $r_p$ in a disk with $(h/r)_p = 0.07$ in transition from the isothermal limit to the adiabaitc limit.
    \item For $Q\gtrsim 5$ disks, the effect of radiative cooling is more critical than self-gravity.
\end{itemize}

Our simulations also have  implications for the observations. We have run planet-disk simulations with massive planets ($\sim$ thermal mass):
\begin{itemize}
    \item One might overestimate the planet mass using the simulation that neglects self-gravity. The gap width is less sensitive to the disk self-gravity than the gap depth. Thus, it is better to constrain the planet mass using the gap width when disk self-gravity is important.
    \item One might underestimate the planet mass using the simulation that neglects the radiative cooling. This is especially the case when $T_{cool}\sim$ 1.
    \item For deep primary gaps in the high planet mass regime, their shapes are less affected by the cooling time, but the secondary gap's position and depth for the gas depend on the cooling time. On the other hand, the secondary gap in dust or brightness temperature is less sensitive to the cooling time since dust drift fast to the primary gap edge forming secondary gaps. The tertiary gap becomes unnoticeable in both gas and dust when $T_{cool}$ = 1.
    \item The dependence of the gap properties (e.g., gap depth, width, and secondary/tertiary gap's position and depth) on the cooling timescale provides a new way to constrain the gaseous disk surface density.
    \item Assuming the gaps at 24 au, 35 au, and 61 au in AS 209 are all associated with the planet at 99 au, the gas surface density at $\sim$100 au in AS 209 should be $\lesssim$ 20 $\mathrm{g\ cm^{-2}}$ using the $T_{cool}\lesssim$ 0.1 cooling constraints, which is consistent with the surface density constraint from $Q \gtrsim$ 1. 
    
\end{itemize}

\section*{Acknowledgements}

SZ and ZZ thank Chao-Chin Yang, Steve Lubow and Jeffery Fung for helpful discussions. ZZ acknowledges support from the National Aeronautics and Space Administration through the Astrophysics Theory Program with Grant No. NNX17AK40G,  the National Science Foundation under CAREER grant No. AST-1753168, and Sloan Research Fellowship. Simulations are carried out with the support from the Texas Advanced Computing Center (TACC) at The University of Texas at Austin through XSEDE grant TG-AST130002, and the NASA High-End Computing Program through the NASA Advanced Supercomputing Division at Ames Research Center.

\section*{Software}
{\tt Dusty FARGO-ADSG} \citep{baruteau2008a, baruteau2008b, baruteau2016},
{\tt Athena++} (Stone et al. 2020, in prep),
{\tt Matplotlib} \citep{matplotlib},
{\tt Numpy} \citep{numpy}, 
{\tt Scipy} \citep{scipy}




\bibliographystyle{mnras}







\bsp	
\label{lastpage}
\end{document}